\documentclass[pdftex,twocolumn]{aastex6}

\usepackage{apjfonts,graphicx,graphics}

\usepackage{graphicx}
\usepackage{natbib}
\usepackage{mathptmx}
\usepackage{amsmath}
\usepackage{wasysym}

\usepackage{times}
\usepackage{epsfig}
\usepackage{ulem}

\begin{document}

%
\title{Baryon Acoustic Oscillation detections from the clustering of massive halos and different density region tracers in \emph{TianNu} simulation}
\author{Yu Liu$^{1}$, Yu Liang$^{2}$, Hao-Ran Yu$^{1,4,5,6,7}$, Cheng Zhao$^{2}$, Jian Qin$^{1}$, Tong-Jie Zhang$^{1,3}$}
\affil{$^1$ Department of Astronomy, Beijing Normal University, Beijing 100875, China; tjzhang@bnu.edu.cn\\
    $^2$ Tsinghua Center for Astrophysics (THCA) \& Department of Physics, Tsinghua University, Beijing 100084, China \\
    $^3$ Center for High Energy Physics, Peking University, Beijing 100871, P. R. China\\
    $^{4}$Kavli Institute for Astronomy \& Astrophysics, Peking University, Beijing 100871, China\\
    $^{5}$Canadian Institute for Theoretical Astrophysics, University of Toronto, M5S 3H8, Ontario, Canada\\
    $^{6}$Tsung-Dao Lee Institute, Shanghai Jiao Tong University, Shanghai, 200240, China\\
    $^{7}$Department of Physics and Astronomy, Shanghai Jiao Tong University, Shanghai, 200240, China}
\begin{abstract}

The Baryon Acoustic Oscillations (BAO) refer to the ripples of material density in the Universe. As the most direct density tracers in the universe, galaxies have been commonly used in studies of BAO peak detection. The spatial number density of galaxies, to a certain extent, reflects the distribution of the material density of our Universe. Using galaxies as matter tracers, we can construct more overlapping empty spheres (defined as DT voids in \citealt{2016MNRAS.459.2670Z}; DT is the abbreviation for "Delaunay Triangulation") than the matter tracers, via Delaunay Triangulation technique. We show that their radii excellently reflect the galaxy number density round them, and they can serve as reliable different density region tracers. Using the data from an unprecedented large-scale $N$-body simulation "TianNu", we conduct some fundamental statistical studies and clustering analysis of the DT voids. We discuss in detail the representative features of two-point correlation functions of different DT void populations. We show that the peak, the position of which corresponds to the average radius of data samples, is the most representative feature of the two-point correlation function of the DT voids. In addition, we also construct another voids, the disjoint voids, and investigate their some statistical properties and clustering properties. And we find that the occupied space of all disjoint voids accounts for about $45\%$ of the volume of the simulation box, regardless of the number density of mock galaxies. We also investigate the BAO detections based on different tracers, i.e. mock galaxies, low-density region tracers (void tracers), and high-density region tracers respectively. Our results show that BAO intensities (the heights of BAO peaks) detected by low/high-density region tracers are enhanced significantly compared to the BAO detection by mock galaxies, for the mock galaxy catalogue with the number density of $7.52\times10^{-5}$ $h^3$ Mpc$^{-3}$.

\end{abstract}

\keywords{cosmology: large-scale structure of Universe: Baryon Acoustic Oscillation: statistics-methods: cosmic void}

\section{Introduction}
\label{sec:intro}

In cosmology, baryon acoustic oscillations (BAO) are produced by the competition between gravity and radiation due to the couplings between baryons and photons before the cosmic recombination, which led to the sound waves propagated in the pre-recombination universe. After the epoch of matter-radiation decoupling, the acoustic oscillations are frozen and correspond to a characteristic scale, determined by the comoving sound horizon at the last scattering surface (\citealt{2015ApJ...798...40X}),

\begin{equation}\label{}
     s=\int_0^{t_{rec}} c_s(1+z)dx=\int_{z_{rec}}^{\infty}\frac{c_s dz}{H(z)},
\end{equation}
where $c_s$ is the sound speed, $H(z)$ is the Hubble expansion rate, and $t_{rec}$ and $z_{rec}$ are the recombination time and recombination redshift respectively, and leave imprint on the cosmic microwave background (CMB) as well as the distribution of galaxies in the later universe even today (\citealt{2003ApJ...594..665B}; \citealt{2003ApJ...598..720S}; \citealt{2005NewAR..49..360E}). The BAO measurements are based on the observation of an excess on the two-point correlation function or a series of wiggles on the power spectrum of the matter density fluctuations, corresponding to the acoustic horizon at the epoch of recombination (\citealt{1970ApJ...162..815P}; \citealt{1970Ap&SS...7....3S}; \citealt{1984ApJ...285L..45B}; \citealt{1996ApJ...471..542H}; \citealt{1998ApJ...496..605E}).

Instead of directly measuring the matter density within the survey volume, astronomers, using galaxies as direct tracers, carried out large field-of-view deep galaxy surveys, e.g. 2dF survey (\citealt{2005MNRAS.362..505C}) and Sloan Digital Sky Survey (SDSS-III) Baryon Oscillation Spectroscopic Survey (BOSS, \citealt{2011AJ....142...72E}), to investigate the large-scale structure of the Universe, and analyzed the clustering of galaxies. By calculating the two-point correlation function (\citealt{1993ApJ...412...64L}) of the data, these oscillations have already been detected more than ten years ago in the SDSS DR3 Luminous Red Galaxy sample (\citealt{2005ApJ...633..560E}) and in the 2dF survey (\citealt{2005MNRAS.362..505C}), which confirmed that the sound horizon is $\sim$150 Mpc in today's Universe, consistent with the WMAP results (\citealt{2014A&A...571A...1P}). In fact the BAO feature also has been evidenced in the cosmic microwave background anisotropies (see Refs. \citealt{2003ApJS..148..175S}; \citealt{2011ApJS..192...18K}; \citealt{2013ApJS..208...19H}; \citealt{2014A&A...571A...1P}), and in the distribution of the Lyman alpha forest (see Refs. \citealt{2013A&A...552A..96B}; \citealt{2013JCAP...04..026S}; \citealt{2015A&A...574A..59D}; \citealt{2017A&A...603A..12B}), and more recently by analyzing the clustering of quasars with the low space density (\citealt{2017arXiv170506373A}).

The BAO scale serves as standard ruler (\citealt{2003ApJ...598..720S}; \citealt{2003ApJ...594..665B}), allowing for the measurements of the angular distance $D_A(z)$ in the transverse direction and the Hubble parameter $H(z)$, to constrain cosmological parameters (\citealt{2005ApJ...633..560E}; \citealt{2014MNRAS.439...83A}) and to study the expansion rate in the radial direction (\citealt{2012MNRAS.426.2719R}), which gives access to better understand the nature of the acceleration completely independent from the supernova technique.
For this reason a large number of surveys have included BAO measurements as an integral part of their science goals, such as the 2dFGRS (\citealt{2001MNRAS.328.1039C}), the SDSS (\citealt{2000AJ....120.1579Y}), the WiggleZ (\citealt{2010MNRAS.401.1429D}), the BOSS (\citealt{2011ApJ...728..126W}), the SDSS-IV/eBOSS, the DESI/BigBOSS (\citealt{2011arXiv1106.1706S}), the DES (\citealt{2006astro.ph..9591A}), the LSST (\citealt{2012arXiv1211.0310L}), the J-PAS (\citealt{2014arXiv1403.5237B}), the 4MOST (\citealt{2012SPIE.8446E..0TD}), the EUCLID survey (\citealt{2009arXiv0912.0914L}), the Tianlai project (\citealt{2015IAUGA..2252187C}), or the WFIRST survey\footnote{\href{http://wfirst.gsfc.nasa.gov}{http://wfirst.gsfc.nasa.gov}} etc..

The intricate large-scale structure of the present-day Universe is the result of gravitational growth of tiny density perturbations (random Gaussian fluctuations) and the accompanying tiny velocity perturbations in the primordial Universe, which has been evidenced primarily by those of temperature fluctuations in the cosmic microwave background (\citealt{1992ApJ...396L...1S}, \citealt{2003ApJS..148....1B}, \citealt{2007ApJS..170..377S}). The "seeds" of such fields of primordial Gaussian perturbations are considered being the result of the early inflationary phase of our Universe. Under the combined actions of gravity and cosmic expansion, the initial tiny density perturbations are continually enhanced. With the gravitational instabilities dominating the dynamical evolution of the Universe, the structure formation evolves from a linear to a highly nonlinear regime, and the Universe gradually evolves into salient and pervasive foam-like pattern, known as the "cosmic web", which has got revealed by major redshift survey campaigns and also has been demonstrated by ever larger computer $N$-body simulations.

In this framework, voids are formed in density minima of the primordial Gaussian density field. As matter evacuated from their interiors, voids internal matter density rapidly decreases and then they become the large and low-density regions, which causes their interiors less to be affected by gravity nonlinear gravitational effects compared to other denser structures and thus to be closer to the initial gaussian field of the Universe (cf. e.g. \citealt{1993MNRAS.263..481V}; \citealt{2004MNRAS.350..517S}). For this reason, cosmic voids become particularly sensitive probes of cosmology and thus can be used in a variety of cosmological studies, e.g. the integrated Sachs-Wolfe effect (\citealt{2008ApJ...683L..99G}; \citealt{2014MNRAS.439.2978C}; \citealt{2015MNRAS.454.2804G}; \citealt{2013A&A...556A..51I}; \citealt{2015MNRAS.446.1321H}); weak gravitational lensing (\citealt{2015MNRAS.454.3357C}); dark energy (\citealt{2012MNRAS.426..440B}; \citealt{2007PhRvL..98h1301P}; \citealt{2010MNRAS.403.1392L}; \citealt{2015PhRvD..92h3531P}; \citealt{2012MNRAS.426..440B}; \citealt{2011MNRAS.411.2615L}); modified gravity theories (\citealt{2013MNRAS.431..749C}; \citealt{2015MNRAS.451.4215Z}; \citealt{2015MNRAS.451.1036C}; \citealt{2015MNRAS.450.3319L}; \citealt{2015JCAP...08..028B}; \citealt{2016PhRvD..94j3524A}; \citealt{2009arXiv0911.1829M}; \citealt{2012MNRAS.421.3481L}); the Alcock-Paczy{\'n}ski test (\citealt{1979Natur.281..358A}; \citealt{2012ApJ...761..187S}; \citealt{2012ApJ...754..109L}; \citealt{2014MNRAS.443.2983S}).

Being distinctive and striking features of the cosmic web but without an unequivocal definition, cosmic voids are rather difficult to be found systematically in surveys with considerable disagreement on the precise outline of such a region (see, e.g. \citealt{2006MNRAS.367.1629S}). So voids can be defined in many different ways, sometimes with different shape restriction (e.g. cubic cells in \citealt{1991MNRAS.248..313K} or spheres in \citealt{2006MNRAS.369..335P}), for different purposes.

\begin{figure*}
\begin{minipage}{\textwidth}
\centering
\raisebox{0.1cm}{\includegraphics[width=0.39\columnwidth]{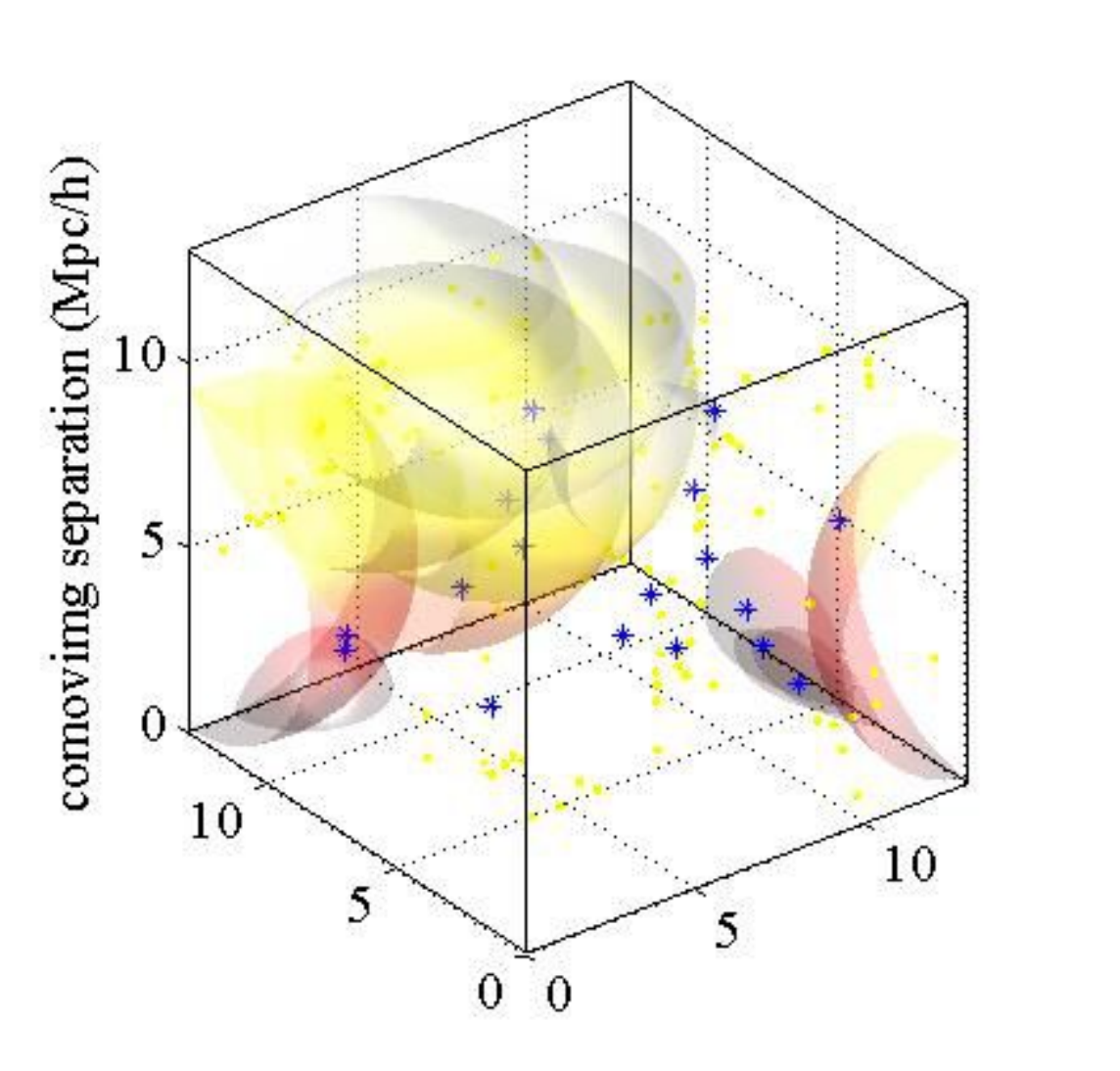}}
\raisebox{0.1cm}{\includegraphics[width=0.39\columnwidth]{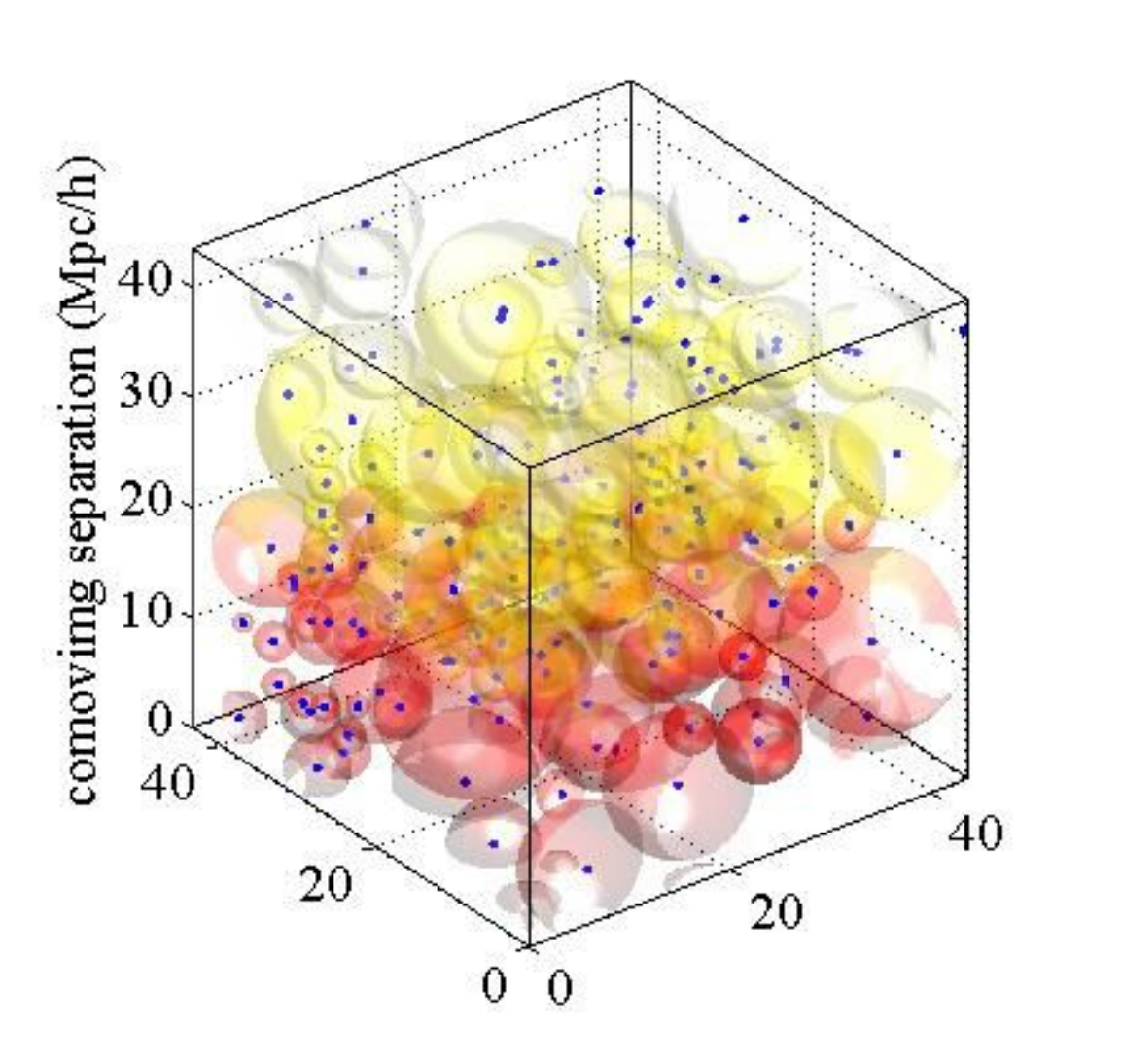}}
\caption{DT voids and disjoint voids. Left panel shows the DT voids which overlap each other severely and cover the entire simulation box. So, in order to better display, we only show a part of the DT voids in the right panel. And the positions of mock galaxies and the positions of DT voids (the centers of the DT spheres) are respectively marked by blue stars and yellow points. Right panel shows another type of voids, the disjoint voids or the non-overlapping voids, which can be straightforwardly obtained from the original DT void catalogue by sorting all DT voids in descending order of radius and then removing the overlapping voids sequently. And the blue points denote the positions of the disjoint voids (the centers of the non-overlapping spheres) in the right panel.
}
\label{fig:showvoid}
\end{minipage}
\end{figure*}

Recently \citealt{2016MNRAS.459.2670Z} developed a $\mathbf{D}$elaunay Tr$\mathbf{I}$angulation (DT, \citealt{delaunay1934sphere}) $\mathbf{V}$oid find$\mathbf{E}$r (\texttt{DIVE}) based on empty circumspheres constrained by tetrahedra of galaxies and named this kind of voids as "\emph{DT voids}". With \texttt{DIVE}, one can construct more DT voids, being the heavily overlapping spheres and covering entire simulation box, than their matter tracers (galaxies/halos), which crucially increases the statistics of void tracers by about 2 orders of magnitude in contrast to previous studies (e.g. Refs. \citealt{2006MNRAS.369..335P}; \citealt{2014PhRvL.112y1302H}; \citealt{2016MNRAS.456.4425C}). In fact, using these overlapping DT voids one can also straightforwardly construct another type of voids, disjoint voids (cf. \citealt{2016MNRAS.459.2670Z}) or non-overlapping voids.

\citealt{2016MNRAS.459.4020L} and \citealt{2016PhRvL.116q1301K} verified that the BAO signal can be clearly detected from these overlapping DT voids, which were separately constructed by mock halo catalogues and by observational data (luminous red galaxies from SDSS-III BOSS DR11). This was not possible previously (see, e.g. \citealt{2006MNRAS.372.1710P}; \citealt{2012ApJ...744...82V}; \citealt{2013MNRAS.431..749C}) for sparse population of voids.

Since the DT voids defined by the Delaunay Triangulation technique are overlapping seriously, we in this work make a definition of the different density regions tracers characterized by the radii of the DT voids, which is actually an indispensable step in reconstruction procedure of DTFE (Delaunay Tessellation Field Estimator) technique. And \citealt{2011ascl.soft05003C} have developed a DTFE public software\footnote{\href{http://www.astro.rug.nl/~voronoi/DTFE/dtfe.html}{http://www.astro.rug.nl/~voronoi/DTFE/dtfe.html}}, which is written in C++ using the CGAL library and is parallelized using OpenMP.

Relying on an unprecedented high-precision $N$-body simulation, "\emph{TianNu}", we in this work perform some fundamental statistical studies and in particular clustering analysis based on the DT voids, and especially study the differences of the BAO signals detected by the mock galaxies (massive halos) and the high/low-density region tracers respectively. Using the density region tracers defined by the DT voids to detect the BAO signals, we find that the BAO peaks of the two-point correlations can be greatly enhanced. Combined with the results from galaxies, it is promising to improve the detection accuracy of the BAO signal, for example, to accurately detect the peak position of the BAO signal.

This paper is organized as follows. First, in \S\ref{sec:simulations} we give a brief introduction of the $N$-body simulation used for this work, and in \S\ref{sec:Mock galaxy} we introduce the Halo Occupation Distribution (HOD) model to construct five mock galaxy catalogues with different number density. In \S\ref{sec:BAO galaxy}, we present the BAO detection using mock galaxies. We present our main analysis results of DT voids and disjoint voids in \S\ref{sec:DT voids} and \S\ref{sec:Disjoint voids}. In \S\ref{sec:void BAO}, we investigate the BAO detections using the different DT void populations, mainly focusing on the BAO detections from the high/low-density region tracers defined by the DT voids. We give a discussion on the differences of the BAO signals detected from different tracers (mock galaxies, low/high-density region tracers) in \S\ref{sec:BAO_tracers difference}. Finally, we summarize and conclude in \S\ref{sec:SUMMARY}.

\section{$N$-body simulations }
\label{sec:simulations}

We ran two large independent $N$-body simulations using the publicly-available high performance cosmological $N$-body code CUBEP3M (\citealt{2013MNRAS.436..540H}; \citealt{2015PhRvD..92b3502I}), one named "\emph{TianZero}" only consisting cold dark matter particles and another one named "\emph{TianNu}" consisting both cold dark matter particles and neutrino particles, both of which are parameterised with [$\Omega_c$, $\Omega_b$, $h$, $n_s$, $\sigma_8$] = [0.27, 0.05, 0.67, 0.96, 0.83] (see \citealt{2017RAA....17...85E}; \citealt{2017NatAs...1E.143Y} for more details). The two simulations, with the same cosmological model and the same seed initial condition and each of them following the evolution of $6912^3$ CDM particles ($13824^3$ neutrino particles are coevolved for massive neutrinos cosmology in one of the two simulations, i.e. TianNu simulation) within a periodic box of size 1.2 Gpc/$h$, respectively took 32 (Tianzero) and 52 (TianNu) hours (11 and 17 million CPU hours) computation time on Tianhe-2\footnote{\href{http://www.nscc-gz.cn/}{http://www.nscc-gz.cn/}}. And the mass resolutions of dark matter particle and neutrino particl are $7\times10^8$ M$_\odot$ and $3\times10^5$ M$_\odot$ respectively.

In this work, our research interests are mainly focused on BAO measurements. Therefore, we only use the data from TianNu to carry out this study. Since no neutrino particles were added in another $N$-body simulation, i.e. TianZero, so, in principle, we can compare these two sets of $N$-body simulation data to study the impact of cosmic massive neutrinos on the formation of large-scale structures of the Universe (e.g. \citealt{2014PhRvD..89j3515U}), such as baryon acoustic oscillations (cf. e.g. \citealt{2015JCAP...07..001P}) and cosmic voids (cf. e.g. \citealt{2015JCAP...11..018M}). We will investigate this in future work.

\section{Mock galaxy catalogues }
\label{sec:Mock galaxy}

 Thanking to such a high-resolution $N$-body simulation, we get the dark matter halo catalogue at redshift z = 0.01, containing 27.758 million dark matter halos corresponding to halo number density of $1.655\times10^{-2}$ $h^3$ Mpc$^{-3}$, by a halo finding procedure using spherical overdensity approach (see \citealt{2017NatAs...1E.143Y} for more details). In the following sections, we will focus our study at redshift z = 0.01 on measurements of the clustering of the five different mock galaxy populations to detect the BAO signals.

In fact, dark matter can not be (unfortunately) directly observed, and usually we rely on large scale structure matter tracers, the observable luminous structures in cosmology (i.e. galaxies and clusters of galaxies), to explore the cosmological information. Indeed, galaxies can be used as tracers of the underlying dark matter distribution to a certain degree, which can be typically described by a multiplicative factor known as the bias. So, we construct five mock galaxy catalogues according to a Halo Occupation Distribution (HOD) model (\citealt{2015JCAP...11..018M}), using the following equation:

\begin{equation}
\left\langle N_c|M\right\rangle=\left\{
             \begin{array}{lr}
             1  & M\geq M_{min},  \\
             0  & M< M_{min},
             \end{array}
\right.
\end{equation}
which means that halos with masses below $M_{min}$ do not host any galaxies, whereas the halo with mass above $M_{min}$ hosts one central galaxy.

The statistical properties of dark matter halos and DT voids (based on dark matter halos as the matter tracers, cf. section \ref{sec:DT voids}) are expected to be sensitive to the number density of haloes/mock galaxies. To investigate this, in our framework, we firstly sort all halos in descending order according to their masses, and then we select out the halos ranking in the top, respectively according to the number density of $7.52\times10^{-5}$ $h^3$ Mpc$^{-3}$, $9.02\times10^{-5}$ $h^3$ Mpc$^{-3}$, $1.05\times10^{-4}$ $h^3$ Mpc$^{-3}$, $1.20\times10^{-4}$ $h^3$ Mpc$^{-3}$, $1.35\times10^{-4}$ $h^3$ Mpc$^{-3}$, to get the five mock galaxy catalogues.

\begin{figure*}
\begin{minipage}{\textwidth}
\raisebox{0.1cm}{\includegraphics[width=0.5\columnwidth]{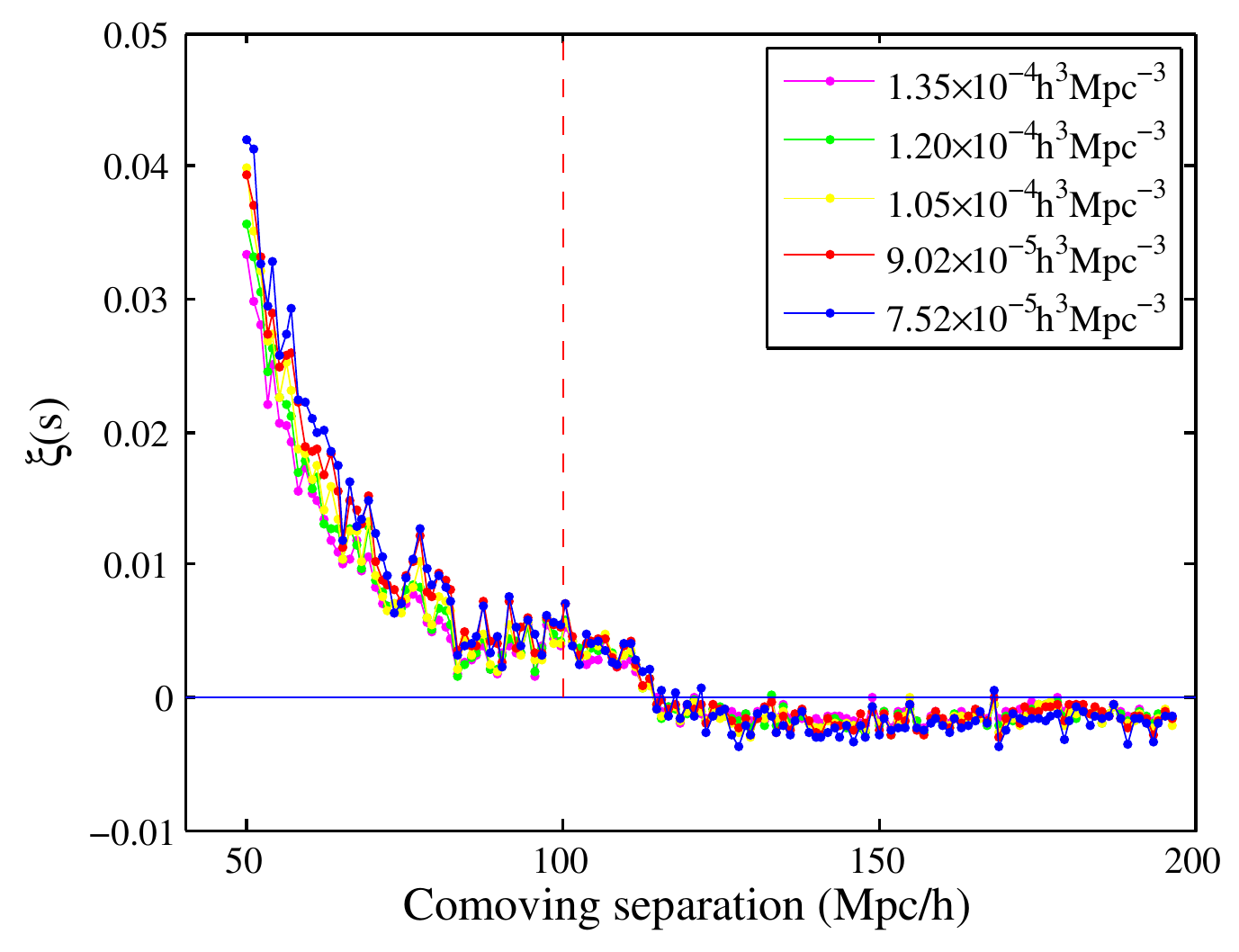}}
\raisebox{0.1cm}{\includegraphics[width=0.5\columnwidth]{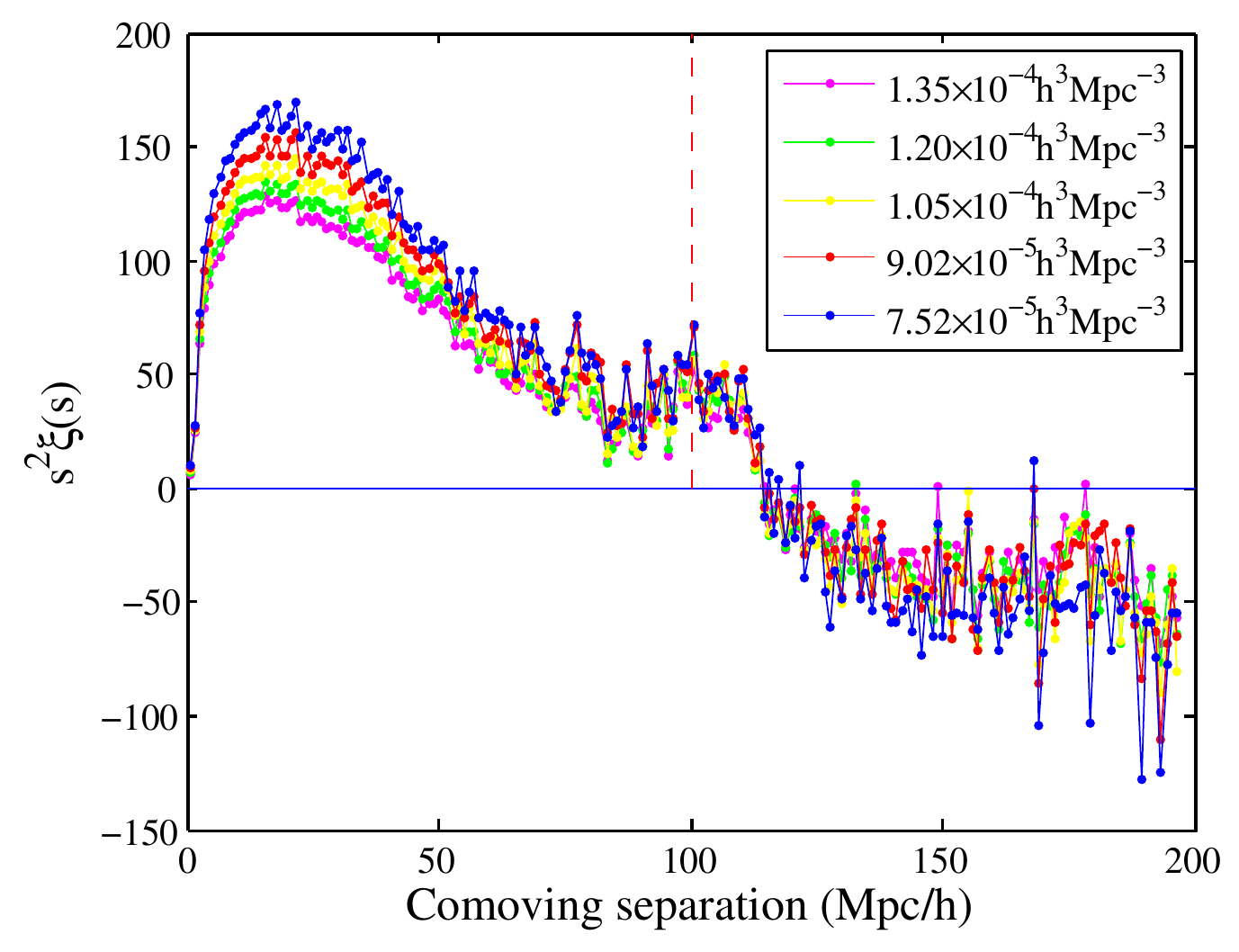}}
\caption{The two-point correlation functions of mock galaxy catalogues. Left panel: the two-point correlation functions, $\xi(s)$, of the five mock galaxy catalogues with the number density of $7.52\times10^{-5}$ $h^3$ Mpc$^{-3}$, $9.02\times10^{-5}$ $h^3$ Mpc$^{-3}$, $1.05\times10^{-4}$ $h^3$ Mpc$^{-3}$, $1.20\times10^{-4}$ $h^3$ Mpc$^{-3}$, $1.35\times10^{-4}$ $h^3$ Mpc$^{-3}$ respectively. Right panel: The two-point correlation functions modulated by the squared distance, $s^2\xi(s)$, to visually enhance the BAO signals. Throughout the following parts of this paper, we will mainly rely on this kind of correlation function plots, $s^2\xi(s)$, to demonstrate the BAO detections.
}
\label{fig:halo}
\end{minipage}
\end{figure*}

\section{BAO signals from Mock galaxy catalogues}
\label{sec:BAO galaxy}

\subsection{The two-point correlation function estimator for mock galaxy catalogues in simulation box }
\label{sec:morph}

The two-point correlation function, commonly used to characterize the large-scale structure of the Universe, has been used to analyze the data of galaxy surveys to describe the probability that one galaxy will be found within a given distance bin of another.

In cosmology, galaxies and dark matter halos, forming in the regions of previous matter overdensities of baryons and dark matter, are mainly configured both at the original sites of the BAO anisotropy and the shells at the sound horizon, which make one would expect to see a greater number of galaxies separated by the sound horizon than by nearby length scales (see \citealt{2005ApJ...633..560E}). Indeed, Baryon Acoustic Oscillations (BAO) measurements are based on the detection of the feature, that corresponds to an excess on the two-point correlation function at a comoving separation equaling to the sound horizon. The BAO signals have been evidenced in many galaxy surveys using galaxies as direct tracers, including the SDSS DR3 Luminous Red Galaxy sample (\citealt{2005ApJ...633..560E}) and the 2dF survey (\citealt{2005MNRAS.362..505C}), where the radial positions of galaxies are measured by their redshifts.

The correlation function $\xi(s)$ ($s$ being the comoving separation) can be computed by different two-point correlation function estimators which have been studied by various authors (\citealt{1974ApJS...28...19P}; \citealt{1983ApJ...267..465D}; \citealt{1982MNRAS.201..867H}; \citealt{1993ApJ...417...19H}; \citealt{2013A&A...554A.131V}). \citealt{2013A&A...554A.131V} have shown that the differences among the two-point correlation function estimators calculated in a cubic geometry are not significant. So, for the convenience of calculation, in our case the correlation function is computed following the \citealt{1974ApJS...28...19P} estimator in the simulation box with periodic boundary condition,

\begin{equation}
\xi(s)=\frac{DD(s)}{RR(s)}-1,
\end{equation}
where the DD (RR) term is the pair count of mock galaxy data samples (random data samples from the Poisson catalogue; cf. \citealt{2000ApJ...535L..13K}; \citealt{1999ApJ...523..480P}) within a given bin of comoving separation [$s-ds/2$, $s+ds/2$] normalized by the total number of pairs.
\subsection{The BAO detection from mock galaxies}
\label{sec:velocity}

 For a more systematic study, we use the five mock galaxy catalogues with the number density of [$2.5\times h^3$; $3\times h^3$; $3.5\times h^3$; $4\times h^3$; $4.5\times h^3$]$\times10^{-4}$ $h^3$ Mpc$^{-3}$ respectively ($h$=0.677) as mentioned in section \ref{sec:Mock galaxy} to carry out our research. This allows us to investigate the impact of the mock galaxy number density on the BAO signal detection for our HOD model, i.e. using massive halos.

 Firstly, we calculate the two-point correlation functions (cf. Fig.\ref{fig:halo}) based on the five mock galaxy catalogues. And then in top panel of Fig.\ref{fig:BAO_tracers} we also show the two-point correlation function curves obtained by Gaussian Process Regression (GPR) fitting (\citealt{2006gpml.book.....R}) of the data points from the results of Fig.\ref{fig:halo}, so that we can understand the trends of the curves more clearly. We find that the BAO signal intensity (the height of BAO peak) decreases with the decline in the number density of mock galaxies, but this effect is not very significant. And it also can be found that the two-point correlation functions of the mock galaxies with lower number density are raised at the comoving separation interval of < $\sim120$ Mpc/$h$, whereas after the intersection at this point ($\sim120$ Mpc/$h$) the curves change from the previous uplift to the depression.

\section{Analysis for DT voids}
\label{sec:DT voids}

\begin{figure*}
\begin{minipage}{\textwidth}
\centering
\includegraphics[width=0.3\columnwidth]{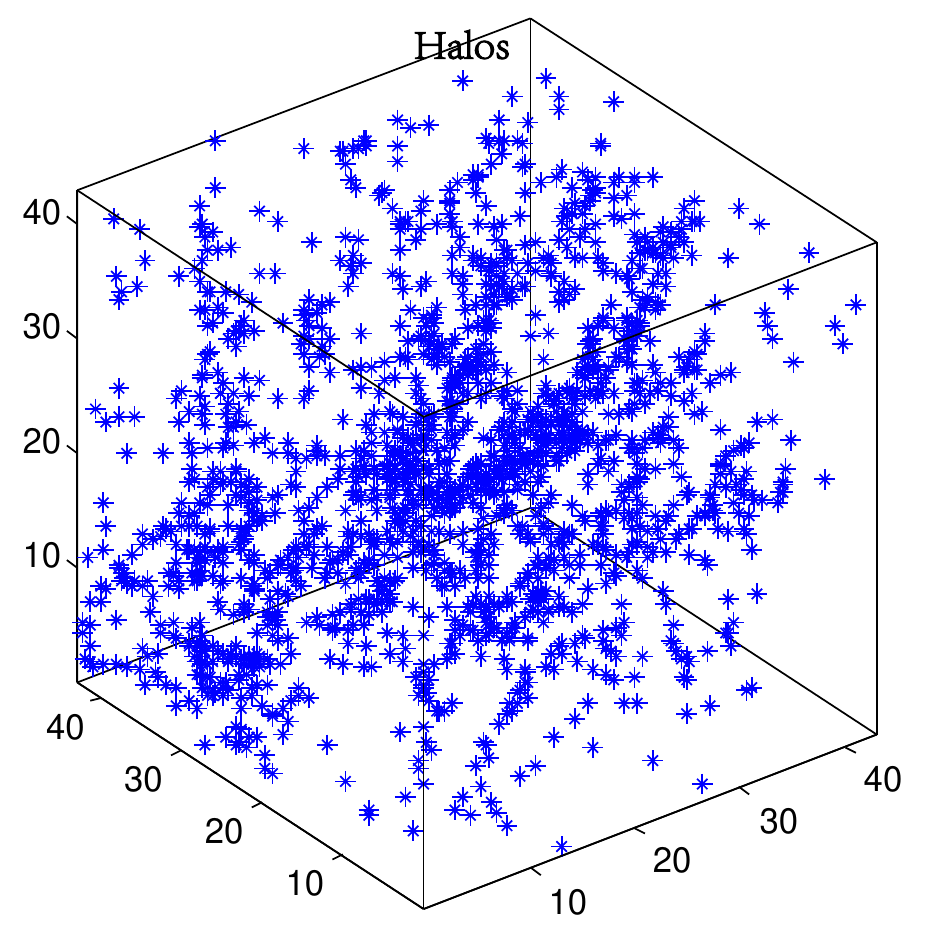}
\includegraphics[width=0.3\columnwidth]{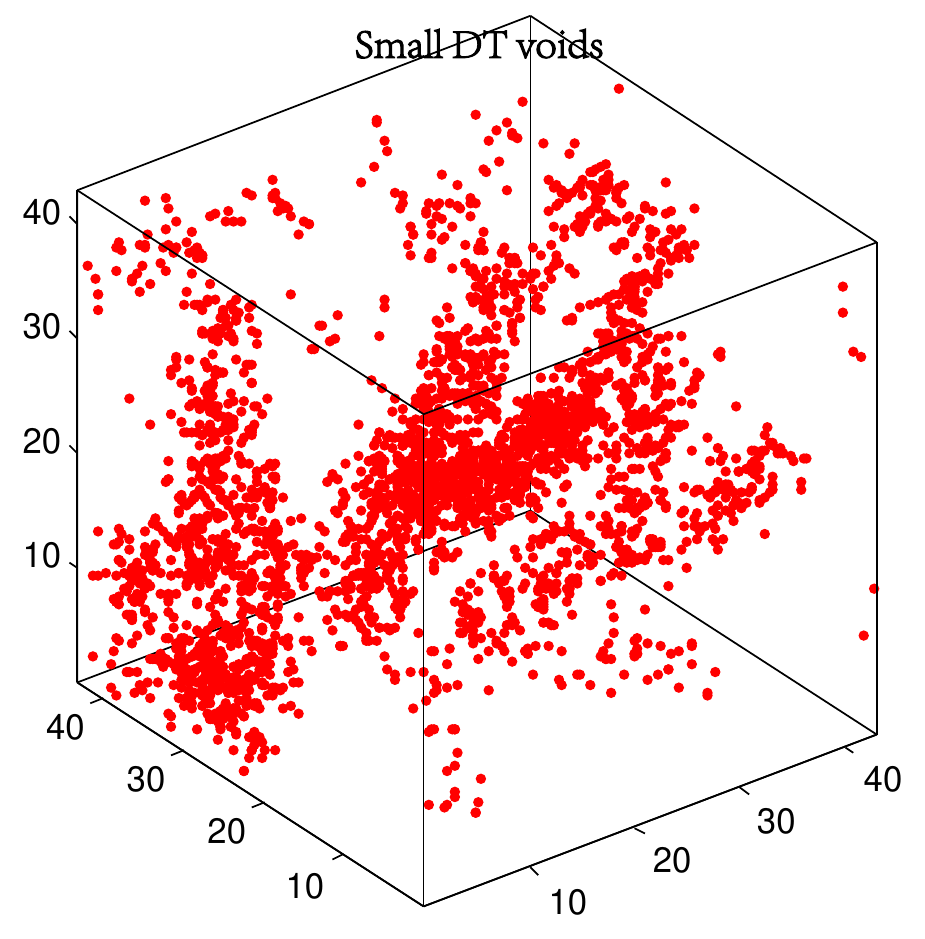}
\includegraphics[width=0.3\columnwidth]{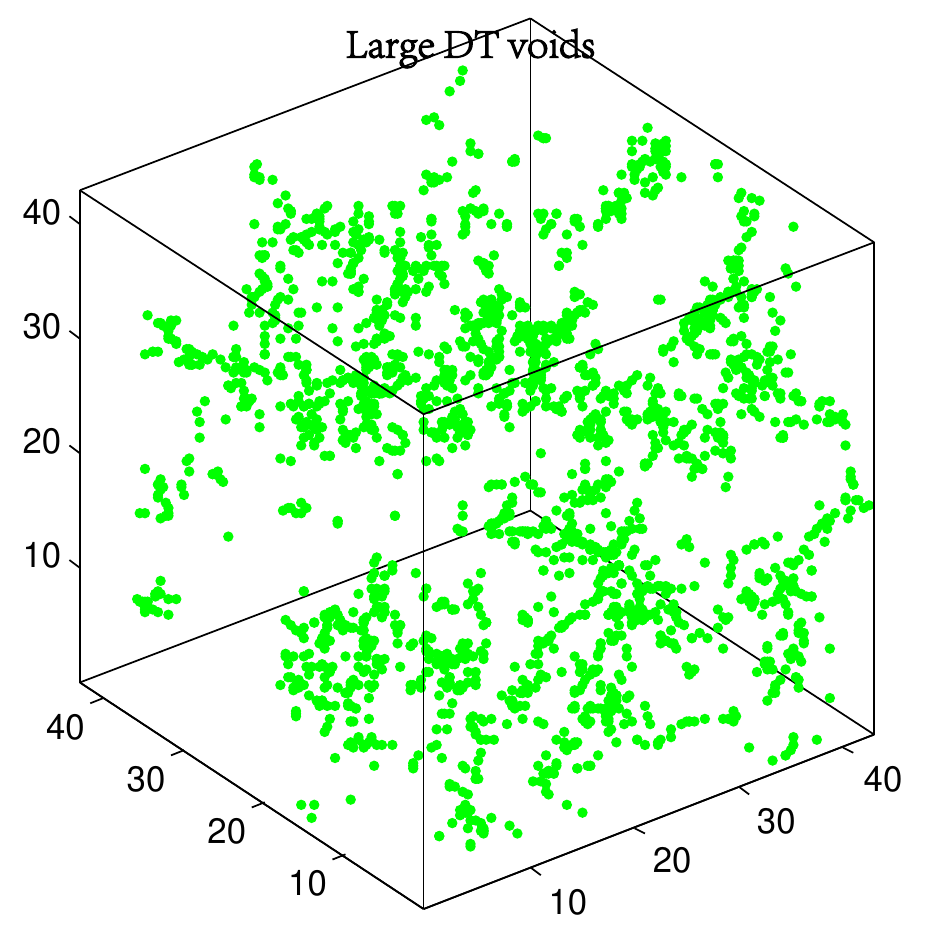}
\includegraphics[width=0.3\columnwidth]{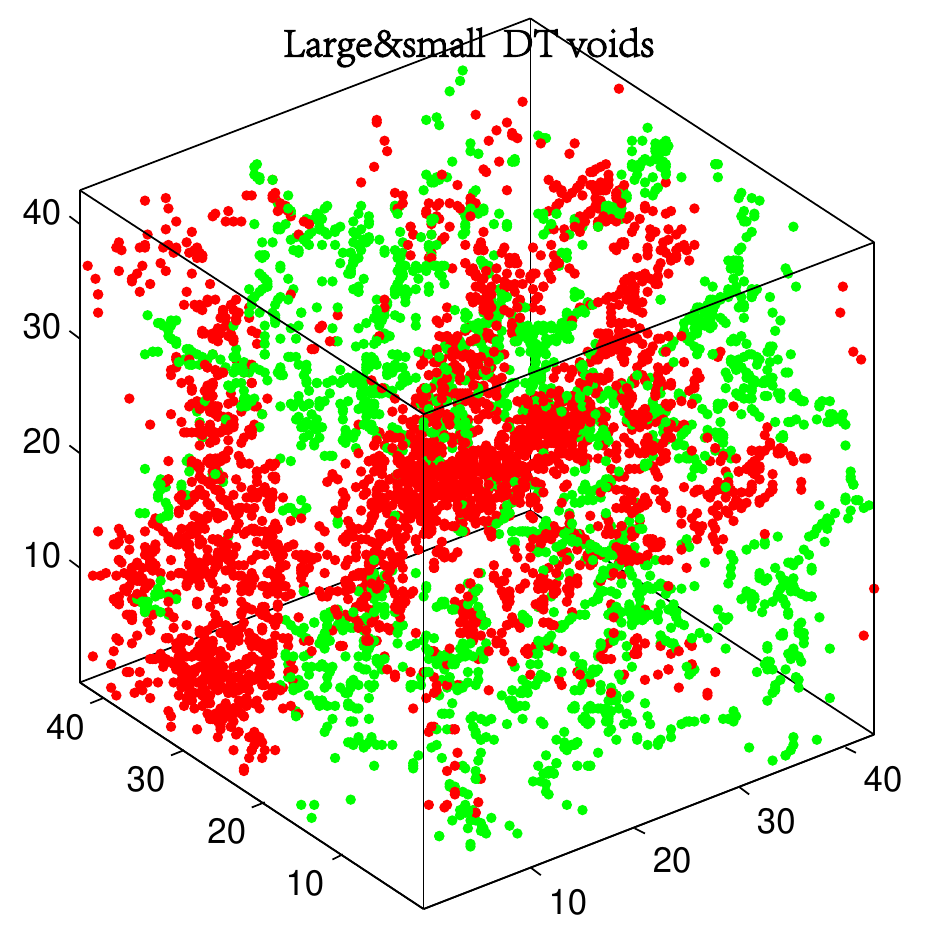}
\includegraphics[width=0.3\columnwidth]{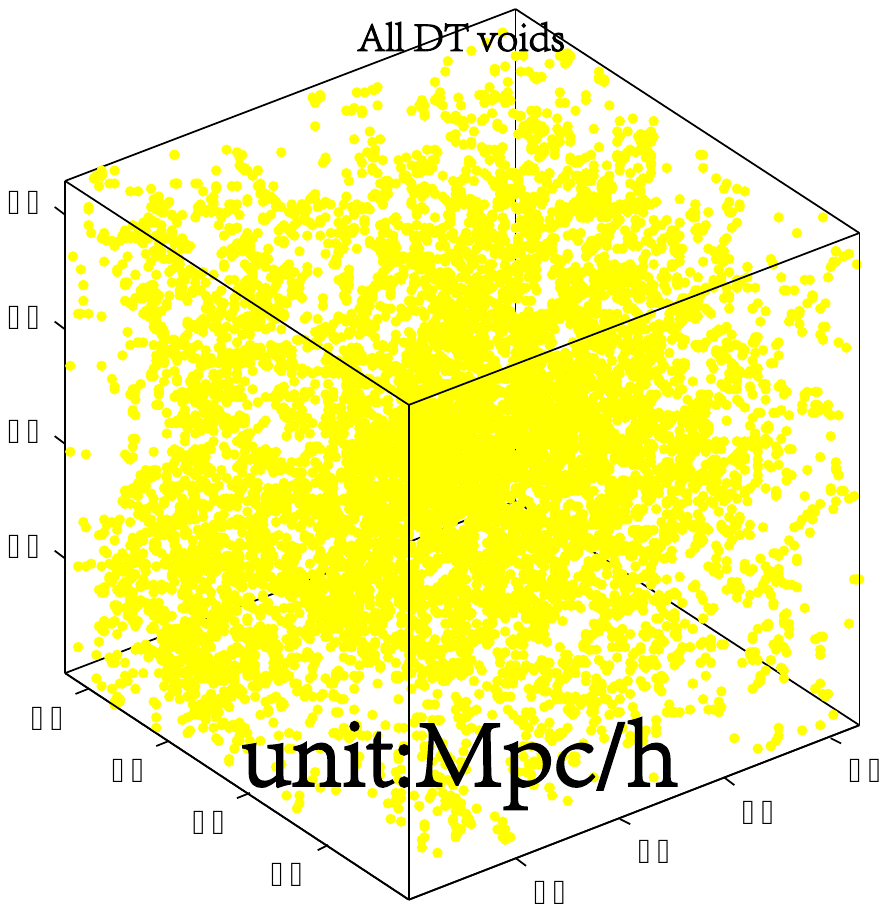}
\includegraphics[width=0.3\columnwidth]{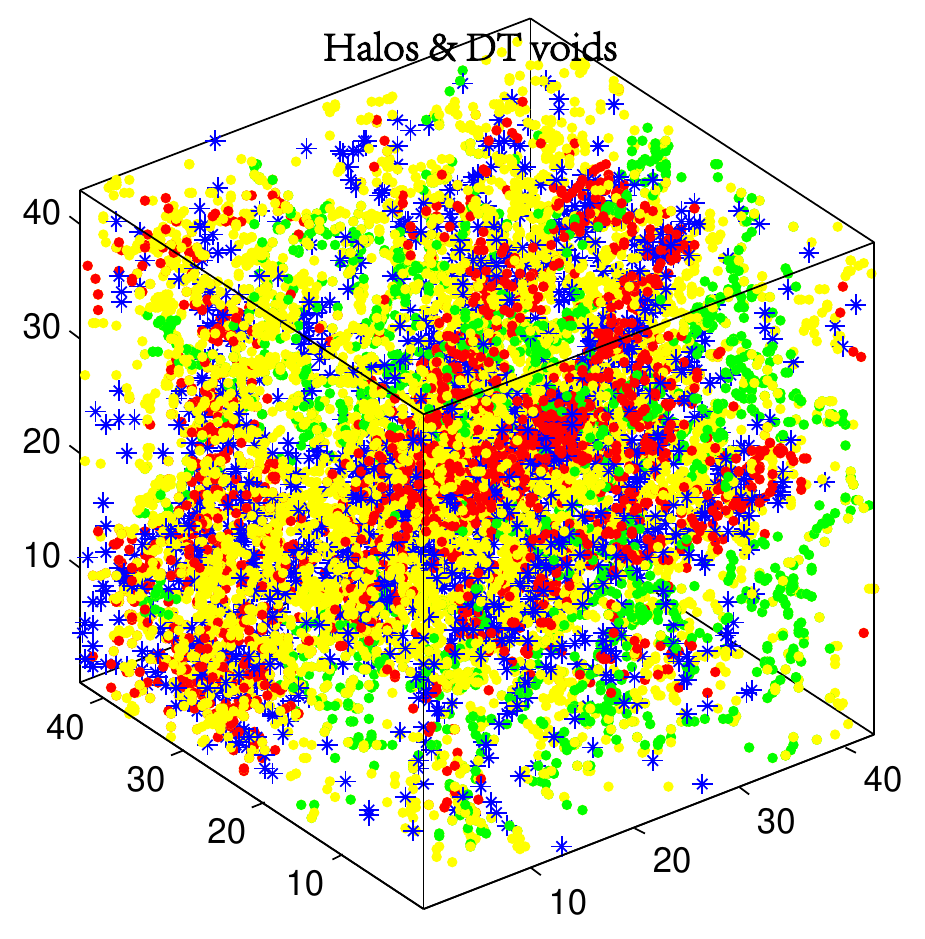}
\caption{The positions of DT voids and mock galaxies (massive halos). The coordinates are in units of $h^{-1}$ Mpc. In the top row, we separately show the positions of the mock galaxies (massive halos), the positions of small DT voids and the positions of large DT voids, which are marked separately by blue stars, red points, green points. And in the bottom row, we separately show the positions of the large\&small DT voids, the positions of all DT voids (with various radii) and the positions of mock galaxies (massive halos) \& all DT voids. It shows the vivid demonstration that the small/large DT voids (high/low-density tracers) well trace the high/low-density regions and all DT voids with various radii tend to be spatially distributed more evenly.}
\label{fig:tracers}
\end{minipage}
\end{figure*}

 Now that we have created five mock galaxy catalogues with different number density, then we run the void-finder \texttt{DIVE} (see \citealt{2016MNRAS.459.2670Z} for more details) on the five mock galaxy fields respectively and construct the five DT void catalogues with the number density of $5.09\times10^{-4}$ $h^3$ Mpc$^{-3}$, $6.11\times10^{-4}$ $h^3$ Mpc$^{-3}$, $7.13\times10^{-4}$ $h^3$ Mpc$^{-3}$, $8.15\times10^{-4}$ $h^3$ Mpc$^{-3}$, $9.16\times10^{-4}$ $h^3$ Mpc$^{-3}$ respectively.

We see that the DT voids of different radii can be used as different density region tracers in section \ref{sec:Density region tracers}. In latter part of this section, we perform some fundamental statistical studies on the DT voids and explore the BAO detections by calculating the two-point correlation functions of different DT void populations within different radius intervals.

\subsection{Density region tracers defined by DT voids of different radii }
\label{sec:Density region tracers}

\begin{figure*}
 \centering
\begin{minipage}{\textwidth}
  \includegraphics[width=0.33\columnwidth]{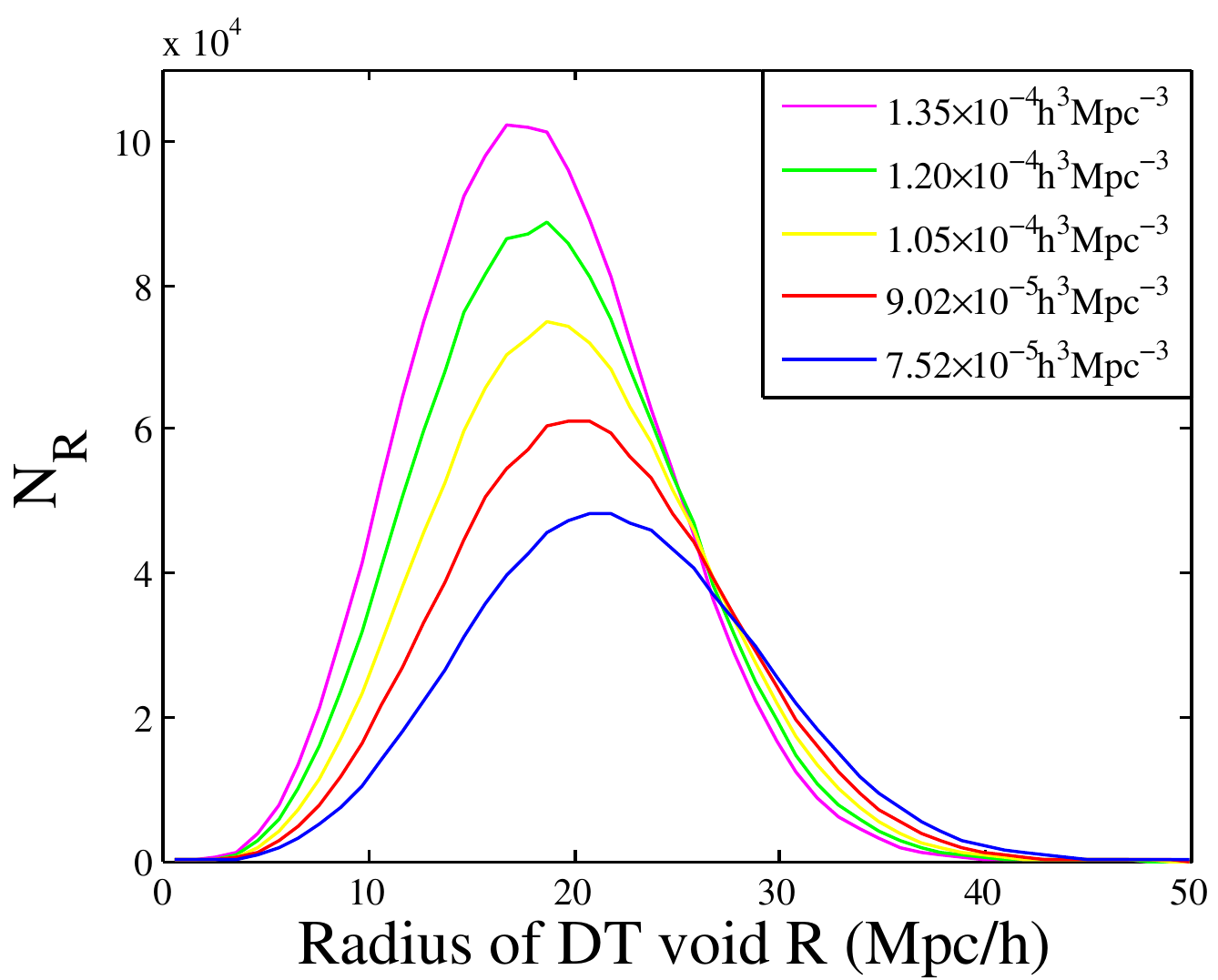}
  \includegraphics[width=0.33\columnwidth]{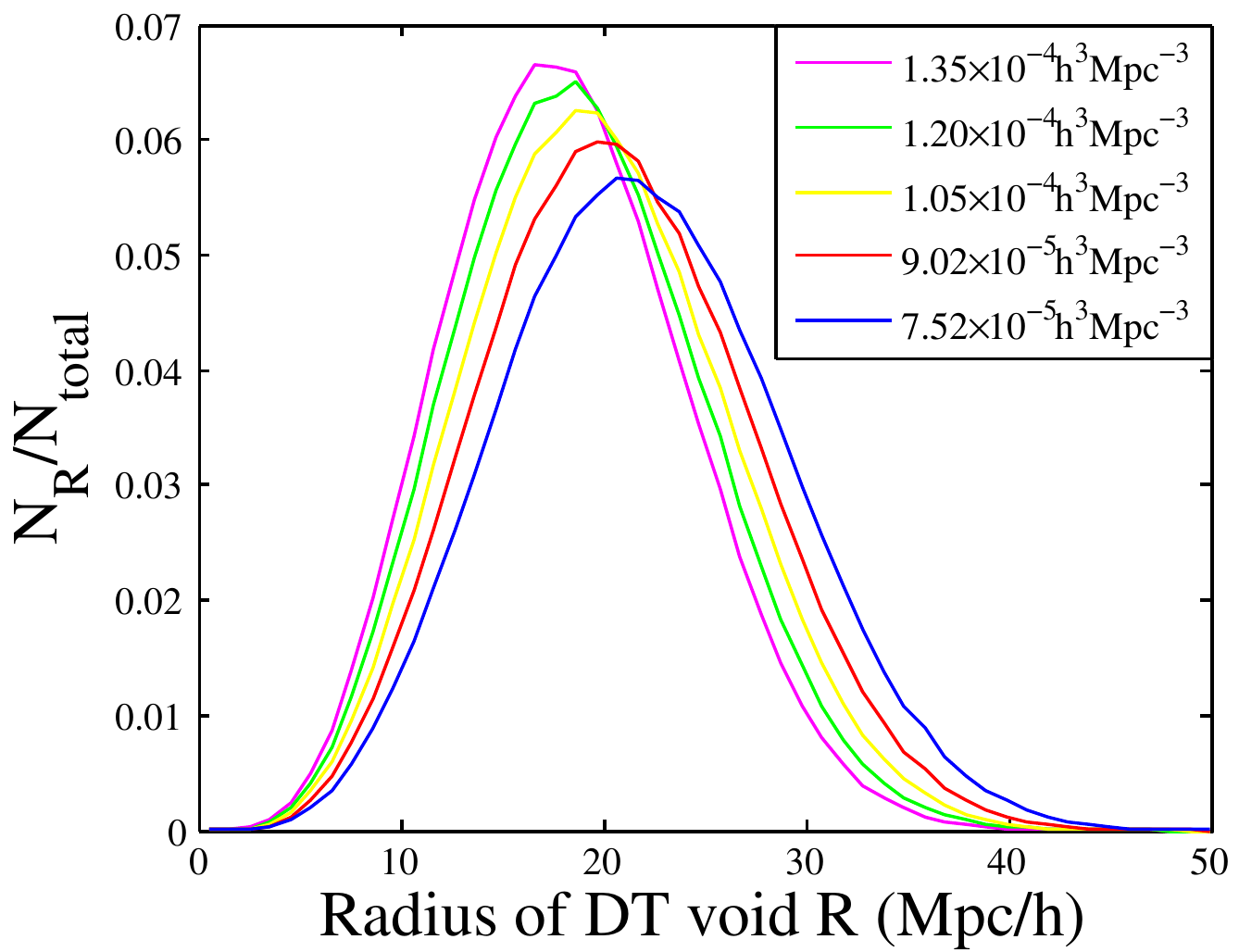}
  \includegraphics[width=0.33\columnwidth]{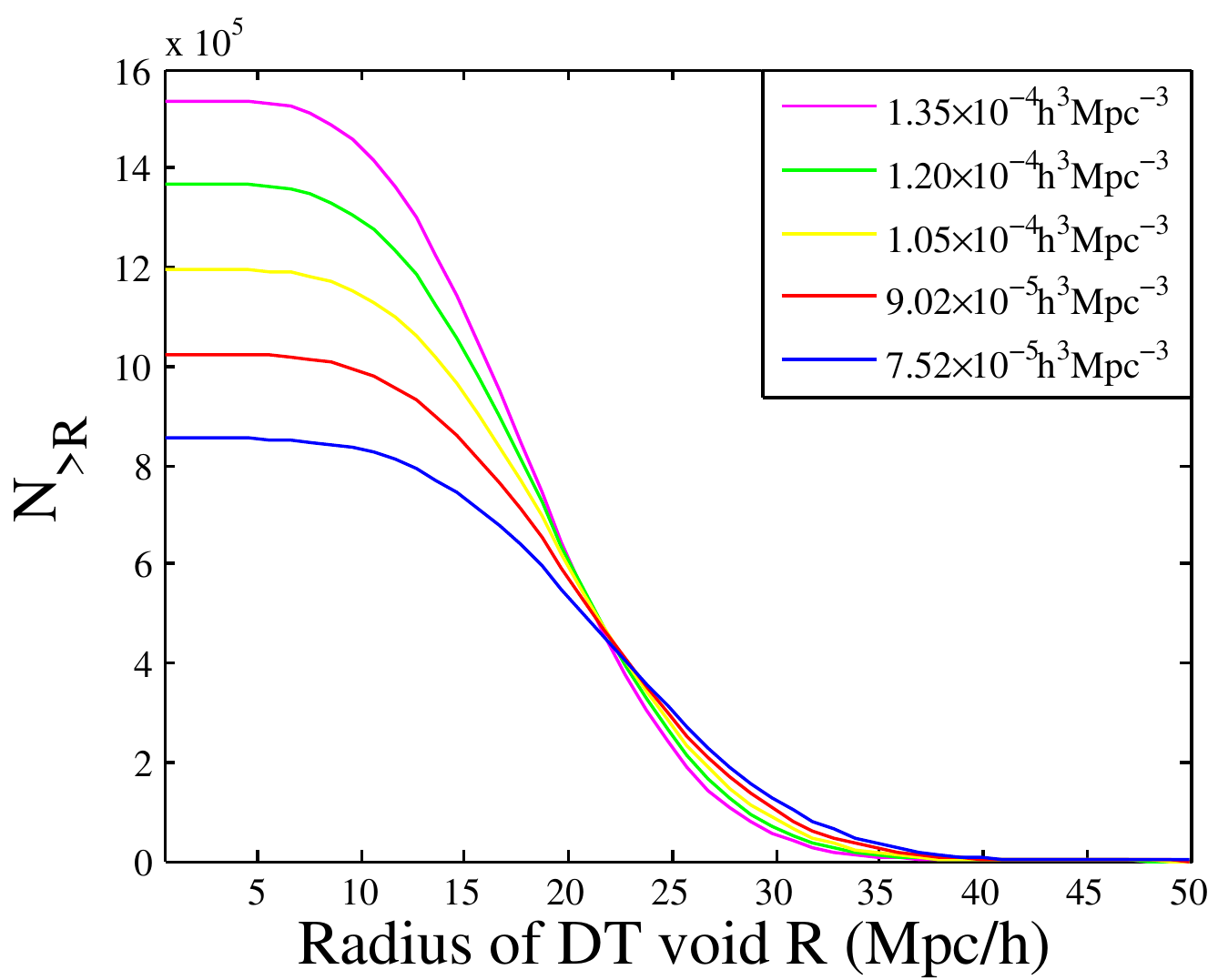}
  \caption{The number functions (or abundances) (left panel), the normalized number functions (PDF) (middle panel) and the cumulative number functions (right panel) with respect to the void radii for the five complete DT void catalogues. The five DT void catalogues are constructed from the five mock galaxy catalogues with the number density of $7.52\times10^{-5}$ $h^3$ Mpc$^{-3}$, $9.02\times10^{-5}$ $h^3$ Mpc$^{-3}$, $1.05\times10^{-4}$ $h^3$ Mpc$^{-3}$, $1.20\times10^{-4}$ $h^3$ Mpc$^{-3}$, $1.35\times10^{-4}$ $h^3$ Mpc$^{-3}$ respectively.}
 \label{fig:numberfunction}
 \end{minipage}
\end{figure*}

The DT voids are the empty circumspheres which are constrained by tetrahedra of galaxies based on the Delaunay Triangulation technique using CGAL\footnote{\href{http://www.cgal.org}{http://www.cgal.org}} (a C++ library of algorithms and data structures for computational geometry) (cf. \citealt{2016MNRAS.459.2670Z}). From this definition, it is not difficult to realize that there are two different populations of DT voids corresponding to their radii. In \citealt{2016MNRAS.459.2670Z} and \citealt{2016MNRAS.459.4020L}, they take small voids as \emph{voids-in-clouds} type voids which have a high probability of residing in dense regions, whereas take large voids as \emph{voids-in-voids} type voids which are more likely to trace underdense expanding regions.

Delaunay Tessellation Field Estimator (DTFE), based on DT technique, represents a natural method of reconstructing a continuous density field from a discrete set of samples (cf. \citealt{2000A&A...363L..29S}; \citealt{2009LNP...665..291V}; \citealt{2011ascl.soft05003C}). The definition of DT voids is actually one of the steps in DTFE (Delaunay Tessellation Field Estimator) reconstruction procedure and the DT voids with different radii definitely trace different density regions, which makes DT voids not as the voids traditionally defined in other literatures. The DT voids overlap each other seriously (cf. Fig.\ref{fig:showvoid}) and cover the entire simulation box, which allows us to get considerable quantity of DT void samples to perform clustering analysis of different void populations at the BAO scales (it was not possible previously, e.g. \citealt{2006MNRAS.372.1710P}; \citealt{2012ApJ...744...82V}; \citealt{2013MNRAS.431..749C}).

In this work, by this void definition, the number of DT voids is about 7 times the mock galaxy population, and the total volume of the DT voids exceeds 200 times the volume of the simulation box. In Fig.\ref{fig:tracers}, we demonstratively illustrate that small DT voids definitely trace the high-density regions and the spatial positions of the large DT voids, tracing the low-density regions, are complementary to the spatial positions of the small DT voids. So in this work, for the more precise definitions, we use new terminologies to rename the small DT voids as high-density region tracers, and rename the large DT voids as low-density region tracers (or void tracers used in \citealt{2016PhRvL.116q1301K}).

\subsection{The number function of DT voids}
\label{sec:mass}

The void statistics have been studied by many works (e.g. \citealt{1986PhRvL..56...99P}; \citealt{1990MNRAS.246..608B}; \citealt{1991MNRAS.248..593E}; \citealt{2002ApJ...566..623B}). Furthermore, the statistics on Delaunay voids, which are constructed from tetrahedra of galaxies and impose the circumspheres to be empty, may include higher order information (see \citealt{1979MNRAS.186..145W}), by encoding higher order statistics through Delaunay Triangulation procedure (see \citealt{2016PhRvL.116q1301K}). So, by analyzing these statistical properties of the DT voids, it is expected to help us circumvent more complicated mathematical formalism to extract some vital information to understand the dynamical processes that affect the structure formation of the Universe (\citealt{2016PhRvL.116q1301K}; \citealt{2016MNRAS.459.4020L}; \citealt{2005MNRAS.356.1155C}; \citealt{2004ApJ...605....1G}; \citealt{2005ApJ...620..618H}).
For the DT voids being of different sizes and overlapping seriously, naturally, the primary knowledge of the statistical properties of the DT voids is the information of their size statistics, which is expected to be more sensitive to the cosmological parameters and the nature of dark energy than previous voids studies (\citealt{2009MNRAS.400.1835B}; \citealt{2013MNRAS.434.2167J}; \citealt{2015MNRAS.451.1036C}).

Actually, void probability distribution function and their cumulative void number density can be used to constrain $\sigma_8$ and $\Omega_m h$ (\citealt{2009MNRAS.400.1835B}). So, for investigating these fundamental knowledge of the statistical properties of the DT void sizes, we respectively, in Fig.\ref{fig:numberfunction}, show the results of the number functions (or abundances), the normalized number functions (the number functions normalized by the total number of DT voids) (PDF), and the cumulative number functions, with respect to the void radii for the five sets of the DT void samples. Through  more or less seeming to be Gaussian bell curves, the size distribution (see left/middle panel of the Fig.\ref{fig:numberfunction}) more likely to be log-normal distribution resembling the void size distributions obtained from the Cosmic Void Catalog (CVC) showed in \citealt{2016ApJ...821..110P} and \citealt{2017ApJ...835...69R}. Since in this study we are not aiming at confirming the accurate distribution the DT voids obey, we postpone such a precision study for future work.

More over, we can see that as increase in the number density of mock galaxies, the amount of DT voids increases, and the average radius of DT voids moves in the smaller direction, while we get more small DT voids and less large DT voids. Furthermore, as shown in Fig.\ref{fig:numberfunction}, with the reduction in the number density of the mock galaxies, the distribution curve of the normalized number function tends to be lower and wider, which means that the corresponding radii distribution of the samples tends to be more discrete with larger variance. Our calculation shows the standard deviations are 5.817 Mpc/$h$, 6.009 Mpc/$h$, 6.224 Mpc/$h$, 6.474 Mpc/$h$, 6.800 Mpc/$h$ respectively, for the five DT void catalogues with the number density in descending order.

\begin{figure}
 \centering
  \includegraphics[width=1\linewidth]{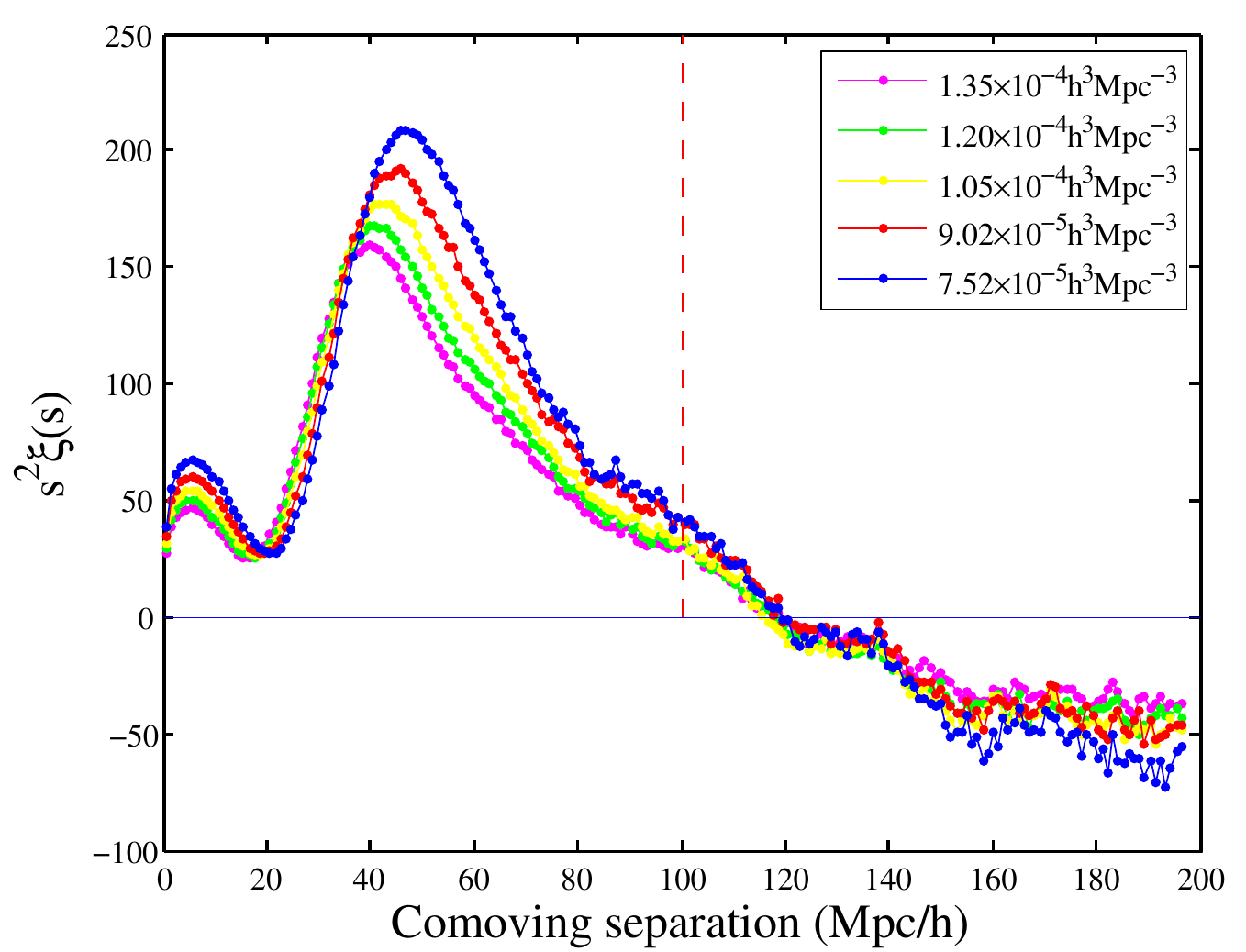}
  \caption{The two-point correlation functions of five complete DT void catalogues with the number density of $5.09\times10^{-4}$ $h^3$ Mpc$^{-3}$, $6.11\times10^{-4}$ $h^3$ Mpc$^{-3}$, $7.13\times10^{-4}$ $h^3$ Mpc$^{-3}$, $8.15\times10^{-4}$ $h^3$ Mpc$^{-3}$, $9.16\times10^{-4}$ $h^3$ Mpc$^{-3}$ respectively. The five DT void catalogues are constructed by the five mock galaxy catalogues with the number density of $7.52\times10^{-5}$ $h^3$ Mpc$^{-3}$, $9.02\times10^{-5}$ $h^3$ Mpc$^{-3}$, $1.05\times10^{-4}$ $h^3$ Mpc$^{-3}$, $1.20\times10^{-4}$ $h^3$ Mpc$^{-3}$, $1.35\times10^{-4}$ $h^3$ Mpc$^{-3}$ respectively.}
 \label{fig:All_DTvoid}
\end{figure}

\subsection{Correlation function from the complete DT void catalogue}
\label{sec:mass}

\begin{figure*}
\begin{minipage}{\textwidth}
\raisebox{0.1cm}{\includegraphics[width=0.5\columnwidth]{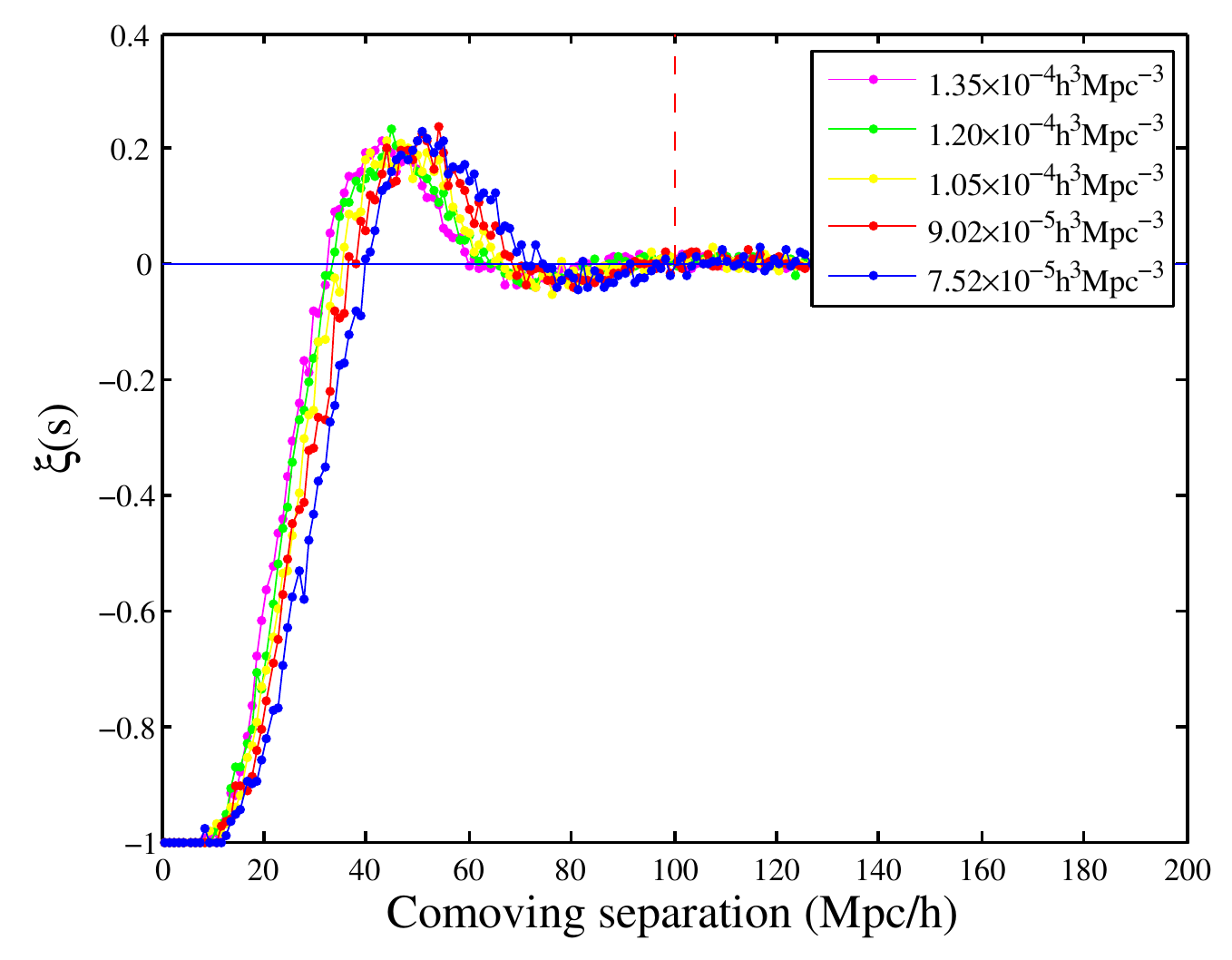}}
\raisebox{0.1cm}{\includegraphics[width=0.5\columnwidth]{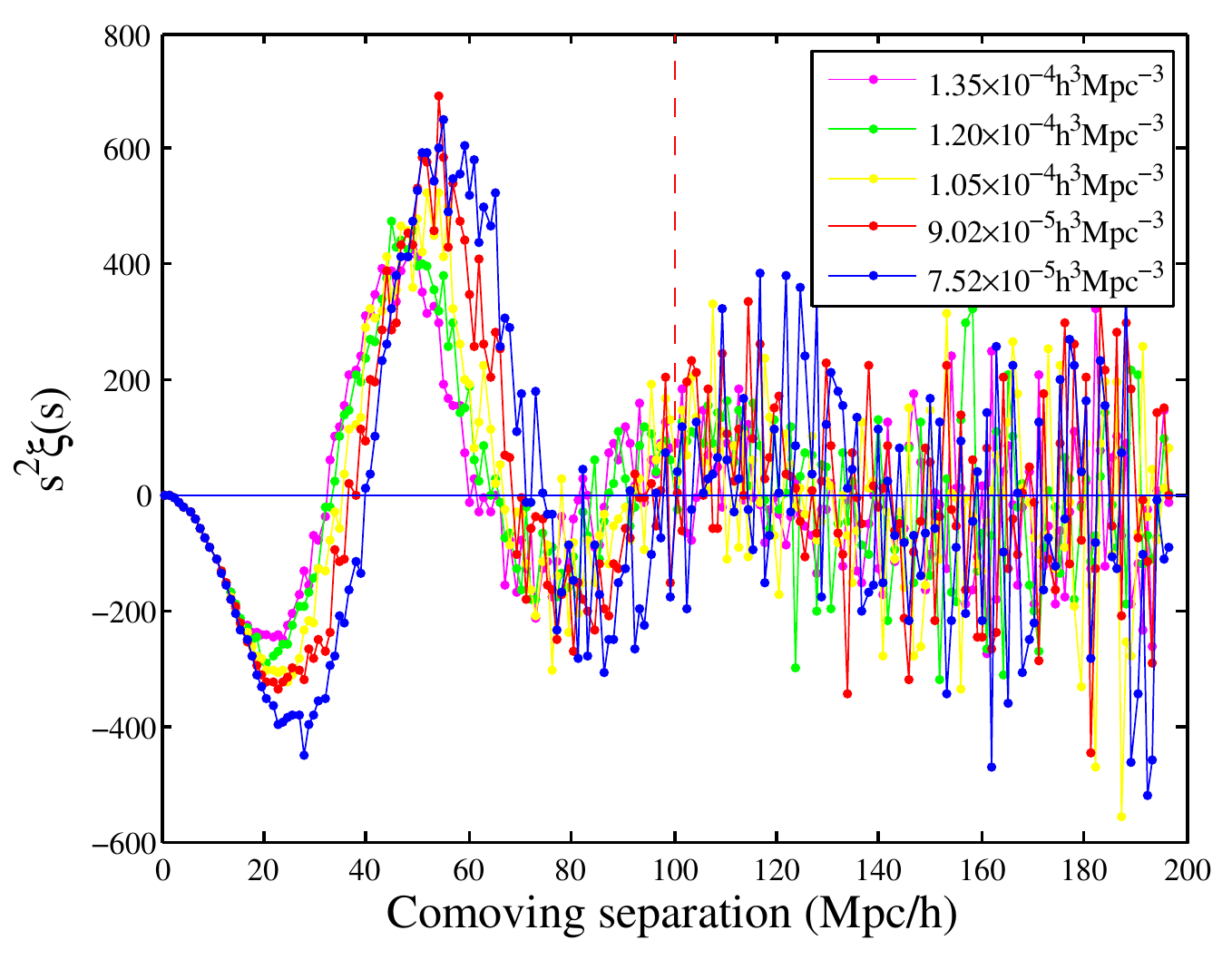}}
\caption{Two-point correlation functions, $\xi(s)$ (left panel), and the two-point correlation functions modulated by the squared distance, $s^2\xi(s)$ (right panel), of the five disjoint void catalogues. The five disjoint void catalogues are constructed by the original five mock galaxy catalogues with number density of $7.52\times10^{-5}$ $h^3$ Mpc$^{-3}$, $9.02\times10^{-5}$ $h^3$ Mpc$^{-3}$, $1.05\times10^{-4}$ $h^3$ Mpc$^{-3}$, $1.20\times10^{-4}$ $h^3$ Mpc$^{-3}$, $1.35\times10^{-4}$ $h^3$ Mpc$^{-3}$ respectively. In the left panel, the obvious void exclusion effect of the disjoint voids are shown clearly, corresponding to the value of the correlation function equals to -1, due to the reason of the non-overlapping pattern. In the right panel, The oscillation patterns of the two-point correlation functions, due to hard sphere exclusion effects with high filling factor (\citealt{1963PhRvL..10..321W}), are not related to the BAO signals, which has been confirmed in \citealt{2016PhRvL.116q1301K}.
}
\label{fig:disjointvoid}
\end{minipage}
\end{figure*}

As mentioned in the introduction, little works focus their study interests on the clustering analysis of cosmic voids for their sparse population and low signal-to-noise ratio. However, there are still some works which have investigated the two-point statistics of the cosmic voids (e.g. Refs. \citealt{1995MNRAS.275.1185G}, \citealt{2005MNRAS.363..977P}, \citealt{2015JCAP...11..018M}, \citealt{2016MNRAS.456.4425C}).

 Galaxies and dark matter halos mainly reside in the high-density regions, for example, Luminous Red Galaxies (LRGs) (cf. \citealt{2015MNRAS.450.1836K}; \citealt{2016PhRvL.116q1301K}). But actually there is also residual matter, which can evolve into galaxies and dark matter halos, in the voids (e.g. \citealt{2011MNRAS.417.1335P}; \citealt{2002A&A...389..405P}; \citealt{2005ApJ...620..618H}; \citealt{2006MNRAS.371..401H}; \citealt{2006MNRAS.372.1710P}). Whereas the complete DT void samples constructed from mock galaxy catalogue have various radii, which means that they locate in different density regions and tend to be evenly distributed in every corner of the entire simulation box (cf. Fig.\ref{fig:showvoid} and Fig.\ref{fig:tracers}). Therefore, if we attempt to use a complete DT void catalogue as a sample set to calculate their two-point correlation function, we expect the BAO signal will not be significant, or simply does not exist.

In order to investigate this, we show the calculation results of the two-point correlation functions of the five complete DT void catalogues in Fig.\ref{fig:All_DTvoid}. As expected, the curves at the BAO scale do not show obvious BAO signals. But we can see a large peak, the position of which moves to the right and the height of which tends to go higher with the decrease in the number density of the mock galaxies. In fact, this peak corresponds to the comoving scale of the average diameter of the DT void samples (cf. Fig.\ref{fig:peakposition}). According to the statistical explanation of the two-point correlation function, that is to say, from the center of a DT void it is most likely to find another center of a DT void at this distance corresponding to the peak position.

\section{Disjoint voids }
\label{sec:Disjoint voids}

 As a matter of fact, because of the vague definitions, there are a plethora of void identification procedures for different purposes or interests. More over, the cosmic voids usually are defined and identified by some void finders under a certain shape restriction (e.g. out of cubic cells in \citealt{1991MNRAS.248..313K} or out of spheres in \citealt{2006MNRAS.369..335P}).

Although the DT voids from a complete catalogue overlap each other seriously, we can straightforwardly obtain another type of voids, disjoint voids (cf. \citealt{2016MNRAS.459.2670Z}) or non-overlapping voids (shown in Fig.\ref{fig:showvoid}), by sorting all DT voids in descending radius order and then removing the overlapping voids sequently. In previous studies, some other non-overlapping types of cosmic voids have been proposed to constraint cosmological parameters (\citealt{2009MNRAS.400.1835B}) and to study galaxy orientation (e.g. \citealt{2006ApJ...640L.111T}; \citealt{2007MNRAS.375..184B}; \citealt{2012ApJ...744...82V}).

Due to the reason of non-overlapping pattern, the two-point correlation functions of disjoint voids show obvious void exclusion effect (\citealt{2014PhRvL.112d1304H}; \citealt{2015JCAP...11..018M}), namely, the probability of finding two non-overlapping void centers at a distance smaller than a certain distance (exclusion scale) is zero, corresponding to that the value of the correlation function equals to -1 (cf. left panel of Fig.\ref{fig:disjointvoid}). The oscillation patterns of their two-point correlation functions (cf. the right panel of Fig.\ref{fig:disjointvoid}), due to hard sphere exclusion effects with high filling factor (cf. \citealt{1963PhRvL..10..321W}), are not related to the BAO signals, which was confirmed in \citealt{2016PhRvL.116q1301K}. In the right panel of Fig.\ref{fig:disjointvoid}, we also find that the curve reaches a maximum, whose height and position depend on the number density and the average size of the disjoint void samples considered (cf. \ref{fig:peakposition}).

\begin{figure*}
\begin{minipage}{\textwidth}
\raisebox{0.1cm}{\includegraphics[width=0.55\columnwidth]{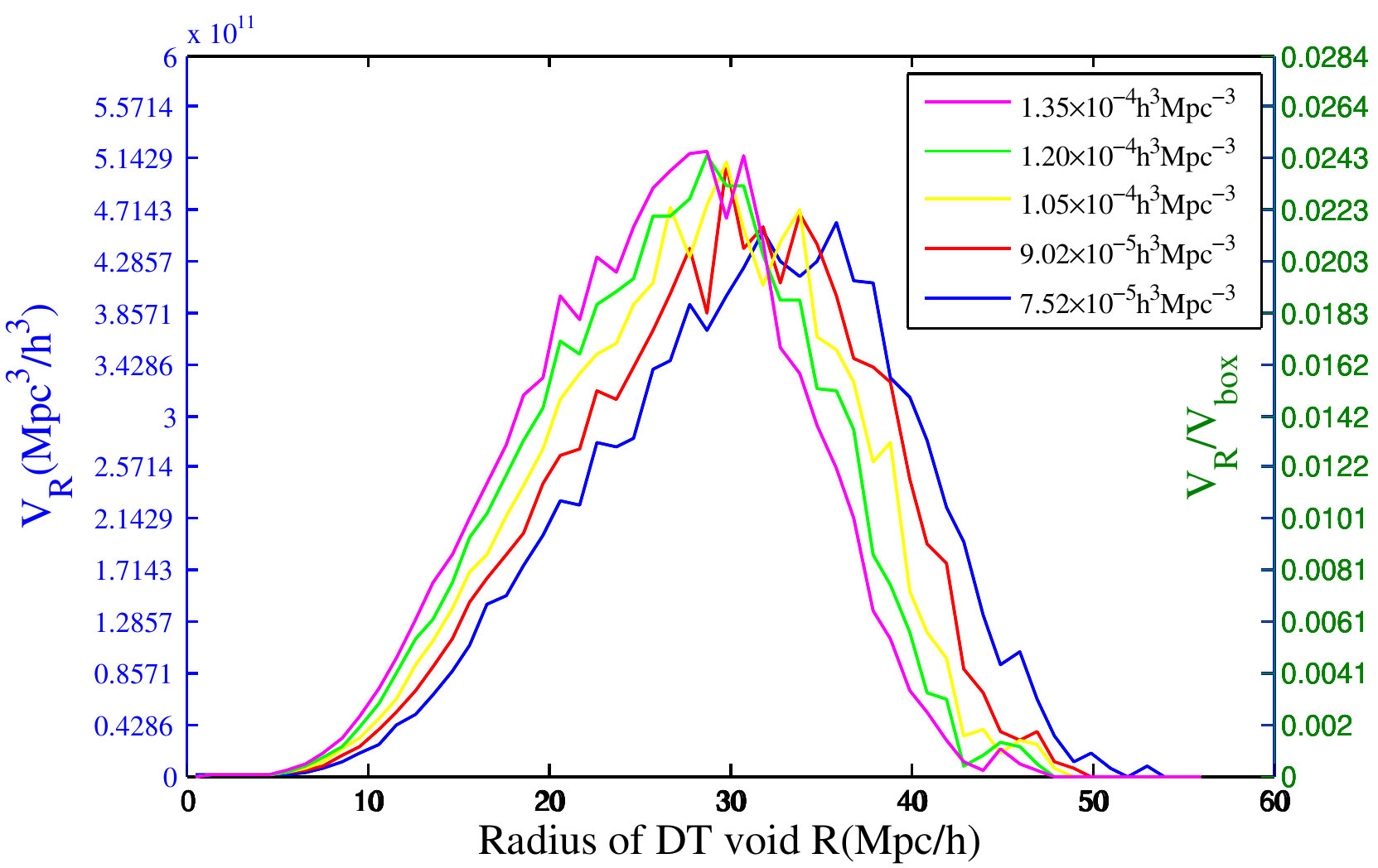}}
\raisebox{0.1cm}{\includegraphics[width=0.437\columnwidth]{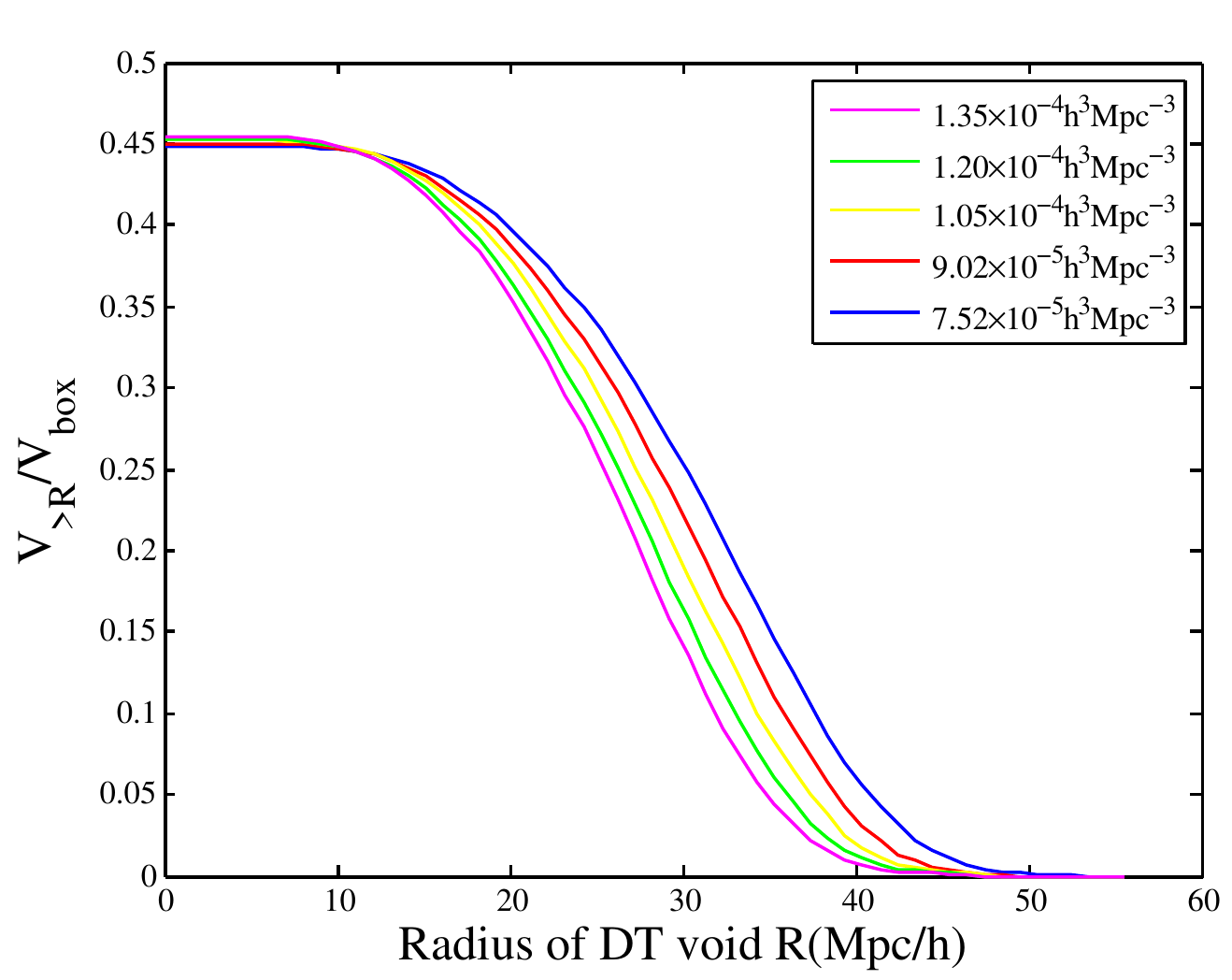}}
\caption{The volume filling fractions (left panel) and cumulative volume filing fractions (right panel) of the five disjoint void catalogues with the number density of $7.59\times10^{-6}$ $h^3$ Mpc$^{-3}$, $9.11\times10^{-6}$ $h^3$ Mpc$^{-3}$, $1.06\times10^{-5}$ $h^3$ Mpc$^{-3}$, $1.21\times10^{-5}$ $h^3$ Mpc$^{-3}$, $1.35\times10^{-5}$ $h^3$ Mpc$^{-3}$ respectively. The five disjoint void catalogues are constructed by the original five mock galaxy catalogues with number density of $7.52\times10^{-5}$ $h^3$ Mpc$^{-3}$, $9.02\times10^{-5}$ $h^3$ Mpc$^{-3}$, $1.05\times10^{-4}$ $h^3$ Mpc$^{-3}$, $1.20\times10^{-4}$ $h^3$ Mpc$^{-3}$, $1.35\times10^{-4}$ $h^3$ Mpc$^{-3}$ respectively. In the left panel, the left ordinate represents  the sum volume of the disjoint voids within a certain radius bin and the right ordinate represents the same one but normalized by the volume of the simulation box.}
\label{fig:volume_fraction}
\end{minipage}
\end{figure*}

In Fig.\ref{fig:volume_fraction}, we show the volume filling fraction and cumulative volume filing fraction of the disjoint voids. As expected, with the decrease in number density of the mock galaxies, the larger disjoint voids account for a larger proportion of the simulation box volume (cf. left panel of Fig.\ref{fig:volume_fraction}), and the volume filling fraction of all disjoint voids decline (cf. right panel of Fig.\ref{fig:volume_fraction}). Interestingly but not surprisingly, although the number density of the mock galaxies changes, the volume filling fraction of all disjoint voids does not change much, about equaling to $45\%$ (cf. \citealt{2016MNRAS.459.2670Z} as a supplement).

\section{BAO detections from different DT void populations with respect to void sizes}
\label{sec:void BAO}

In this section, we investigate the two-point correlation functions of different DT void populations characterized by void sizes. We give a detail discussion on the main features of the DT void two-point correlation functions. And we focus our study on the BAO detections by low/high-density region tracers in the later part of this section. Some of these work was also shown in \citealt{2016MNRAS.459.4020L} based on other halo catalogues of different type (not from $N$-body simulation), which are constructed with the \textbf{P}erturb\textbf{A}tion \textbf{T}heory \textbf{C}atalogue generator of \textbf{H}alo and galax\textbf{Y} (\texttt{PATCHY}; \citealt{2014MNRAS.439L..21K}) distributions. But in our study, first time using unprecedented large-scale $N$-body simulation data, we show some new results and give a more detailed discussion with different purpose.

With our HOD model for constructing the mock galaxy catalogues, we show that even with sparse mock galaxy samples (massive halos), interestingly, we can also detect significant BAO signal, which to a certain degree is not very sensitive to the number density of the mock galaxy samples as showed in Fig.\ref{fig:halo} and in the top panel of Fig.\ref{fig:BAO_tracers}. Here, we use the mock galaxy catalogue with the lowest number density, i.e. $7.5\times10^{-5}$ $h^3$ Mpc$^{-3}$, as our target sample set.

\begin{figure}
 \centering
  \includegraphics[width=1\linewidth]{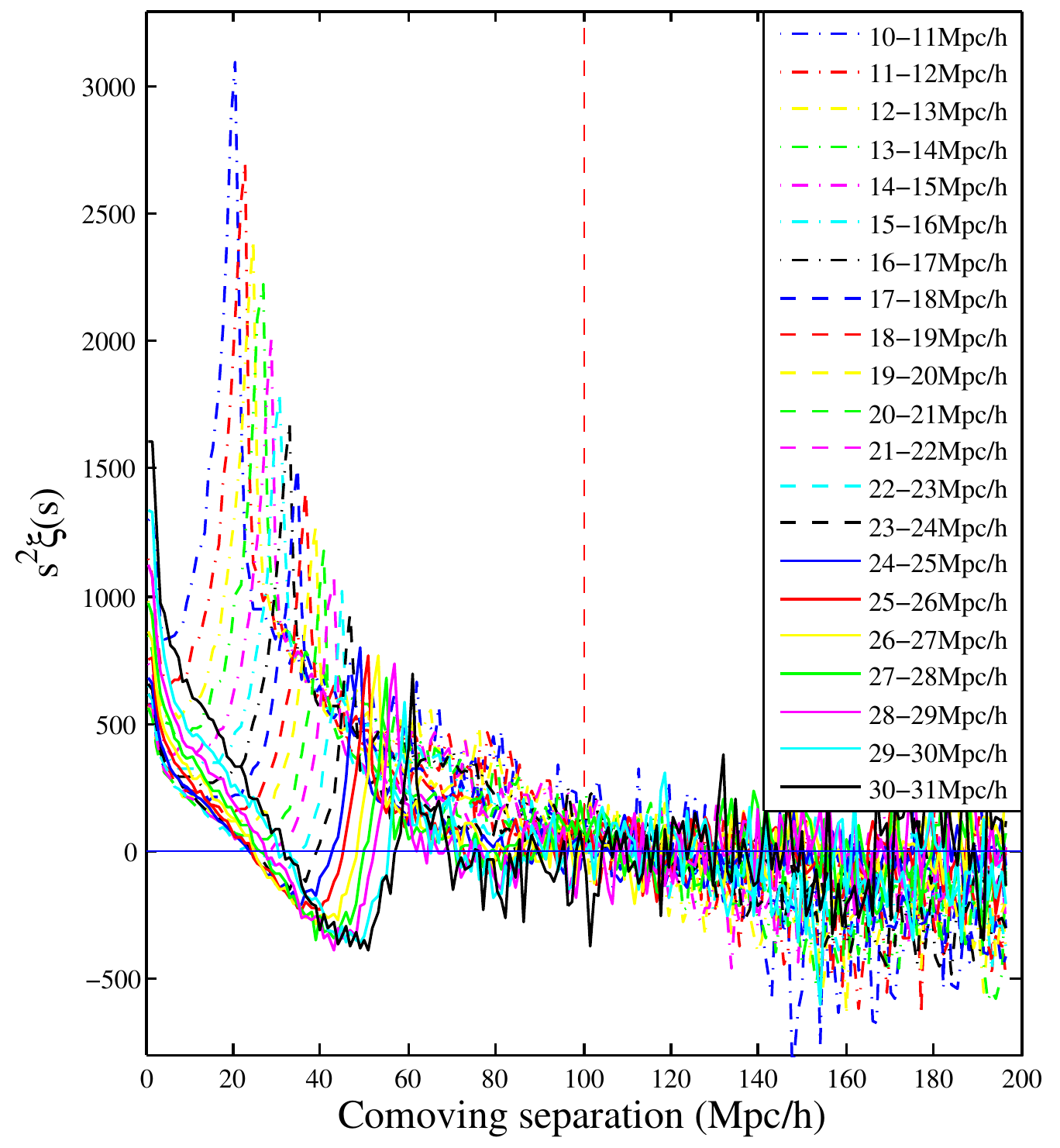}
  \caption{The two-point correlations of the DT voids within different radius bins from the DT void catalogue with the number density of $5.09\times10^{-4}$ $h^3$ Mpc$^{-3}$. The DT void catalogue is constructed by the mock galaxy catalogue with number density of $7.52\times10^{-5}$ $h^3$ Mpc$^{-3}$. The correlation functions show the pronounced and sharp peaks, the locations of which are corresponding to the comoving scales of the average diameters of the data sample sets.
  }
 \label{fig:bin_void}
\end{figure}

\subsection{Correlation functions from different DT void radii bins}
\label{sec:void radii bins}

In section \ref{sec:Density region tracers}, we have already shown that DT voids of different scales well trace different density regions, which makes the DT voids can serve as reliable tracers of different density regions. This encourages us to investigate the typical features of the two-point correlation functions for different DT void populations (i.e. the different density region tracers characterized by DT void radii). So, in Fig.\ref{fig:bin_void} we show the results of the two-point correlation functions of DT void samples, constructed from the mock galaxy catalogue with number density of $7.52\times10^{-5}$ $h^3$ Mpc$^{-3}$, within different radii bins.

As expected, some distinctive features are revealed. Similar to the large peaks in Fig.\ref{fig:All_DTvoid}, pronounced and sharp peaks are shown on these two-point correlation functions. And, in Fig.\ref{fig:peakposition}, we will confirm that these peaks definitely correspond to the average diameter comoving scales of the DT void samples within different radii bins. Surprisingly, regardless of the sample number of the DT voids, the peak height declines smoothly with the increase in the scale of the radii bin. More over, we also find there is a trough on the left side of each peak, which is due to the void exclusion effect (\citealt{2014PhRvL.112d1304H}). As the same time, we can see that for the larger DT void samples, the troughs go deeper until below zero, which means that the larger DT voids tend to reside in low-density regions (i.e. the cosmic void regions) with void exclusion effect becoming more pronounced. Unfortunately, in our case, there are no significant BAO signals shown on these correlation functions.

\subsection{The BAO signals from low/high-density region tracers}
\label{sec:mass}

Galaxies, as the most direct tracers of density in the Universe, have been commonly used in studies of BAO peak detection (e.g. \citealt{2005ApJ...633..560E}; \citealt{2005MNRAS.362..505C}; \citealt{2007MNRAS.381.1053P}; \citealt{2010MNRAS.401.2148P}; \citealt{2014MNRAS.441...24A}; \citealt{2017MNRAS.470.2617A}; \citealt{2011MNRAS.416.3017B}; \citealt{2015MNRAS.449..835R} etc.). And in the former sections, we have already shown in detail that the locations of DT voids can be used as reliable material density tracers. So it is then necessary to explore the BAO detections with different density region tracer populations (i.e. different DT void populations characterized by their radii). Actually, \citealt{2016MNRAS.459.4020L} have investigated and confirmed that for the large DT void samples (low-density region tracers or void tracers), there is an optimal radius cut making the BAO signal best, and there are two dips on both sides of the BAO peak, with this feature of which they also defined a efficient model-independent estimator of S/N (Signal-to-Noise) ratio of BAO signal.

\begin{figure*}
\begin{minipage}{\textwidth}

\raisebox{0.1cm}{\includegraphics[width=1\columnwidth]{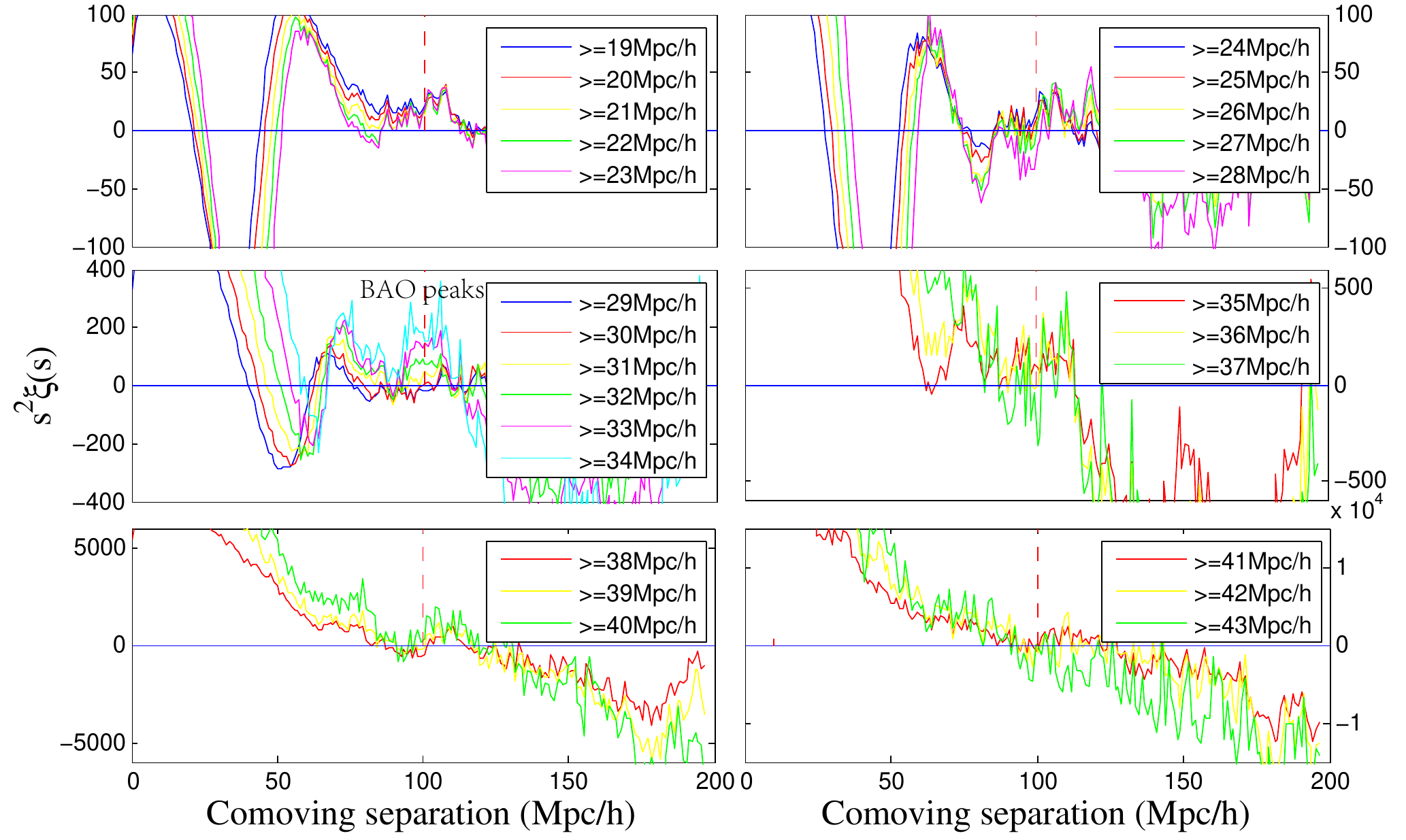}}
\caption{The two-point correlation functions of DT voids with various radius cuts ($R > R_{cut}$) from the DT void catalogue with number density of  $5.09\times10^{-4}$ $h^3$ Mpc$^{-3}$. The DT void catalogue is constructed from the mock galaxy catalogue with number density of $7.52\times10^{-5}$ $h^3$ Mpc$^{-3}$. It confirms that there is a optimal void radius cut (here it is $\sim$34 Mpc/$h$) for the strongest BAO signal intensity from a certain complete DT void catalogue.}
\label{fig:low_density_tracers}
\end{minipage}
\end{figure*}

\begin{figure}
 \centering
  \includegraphics[width=1\linewidth]{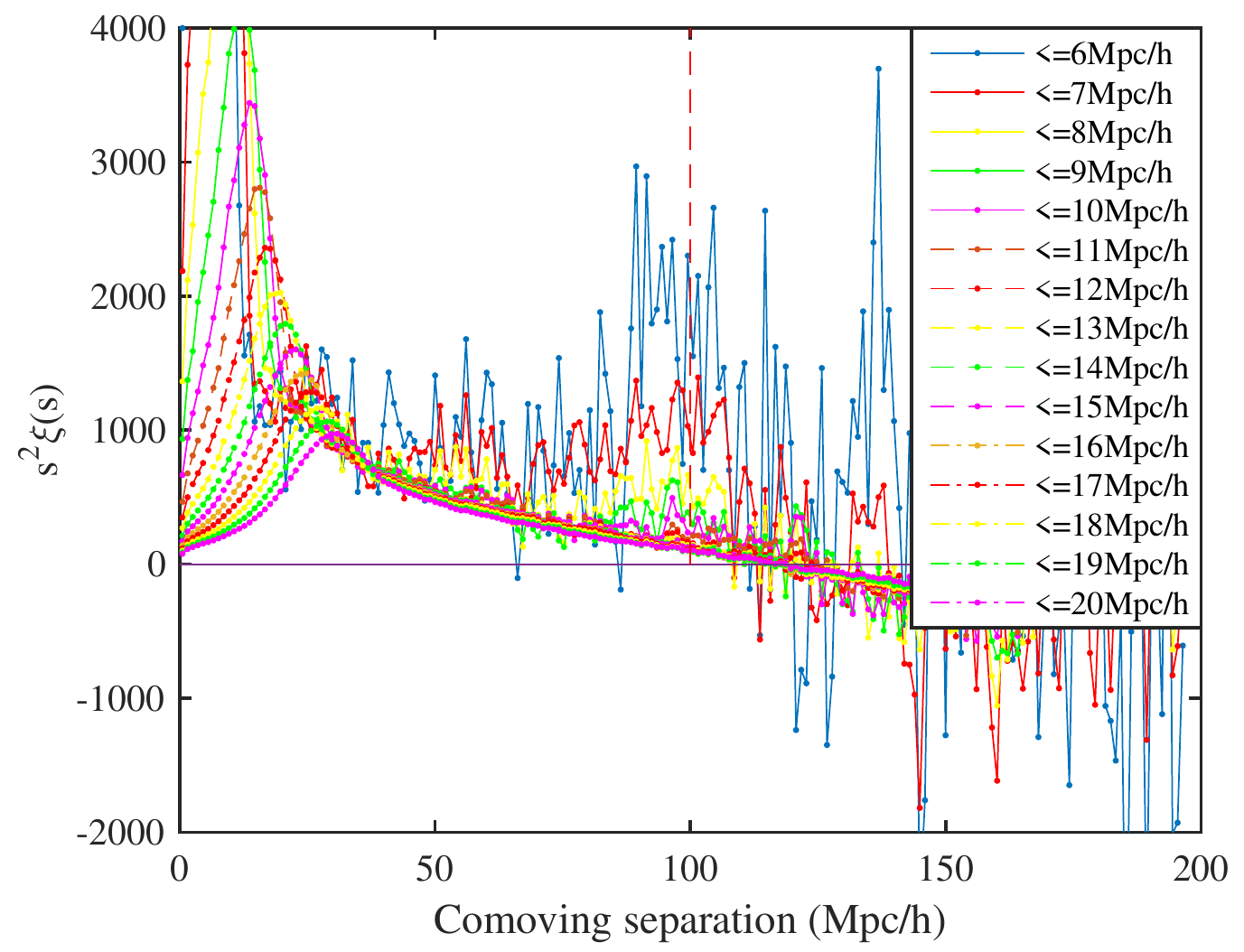}
  \caption{The two-point correlation functions of DT voids with various radius cuts ($R < R_{cut}$) from the DT void catalogue with number density of  $5.09\times10^{-4}$ $h^3$ Mpc$^{-3}$. The catalogue is constructed from the mock galaxy catalogue with number density of $7.52\times10^{-5}$ $h^3$ Mpc$^{-3}$. For lower radius cuts, the two-point correlation functions show stronger noises due to sparse populations.}
  \label{fig:BAO}
\end{figure}

In this section, we mainly investigate the BAO detections with the low/high-density region tracers, and discuss the difference of the features of their two-point correlation functions and the BAO detections by the two methods. To this end, we show the results of the two-point correlation functions of the low/high-density region tracer populations, constructed by the mock galaxy catalogue with number density of $7.52\times10^{-5}$ $h^3$ Mpc$^{-3}$, with different radius cuts, in Fig.\ref{fig:low_density_tracers} and Fig.\ref{fig:BAO} respectively.

We also confirm that there definitely is a optimal radius cut, $\sim34$ Mpc/$h$ in this case, for the best BAO detection based on large DT void populations (low-density region tracers), using our $N$-body simulation data. But for the high-density region tracer populations, it shows the different scenario, where as decline in the radius cut, the BAO peak rise and go higher without optimal radius cut, and as the same time the noise becomes more pronounced due to fewer DT void samples.

It's interesting to note that the two main features discussed in section \ref{sec:void radii bins} are maintained on the two-point correlation functions of the low-density region tracer populations, i.e. the peak corresponding to the average diameter of DT void samples and the trough due to the void exclusion effect; but for the high-density region tracer populations, there is the peak without the trough, which means that they are located in the dense mock galaxy regions. This also unveil the fact that this peak is the most representative feature of the two-point correlation function of the DT voids.

\begin{figure}
 \centering
  \includegraphics[width=1\linewidth]{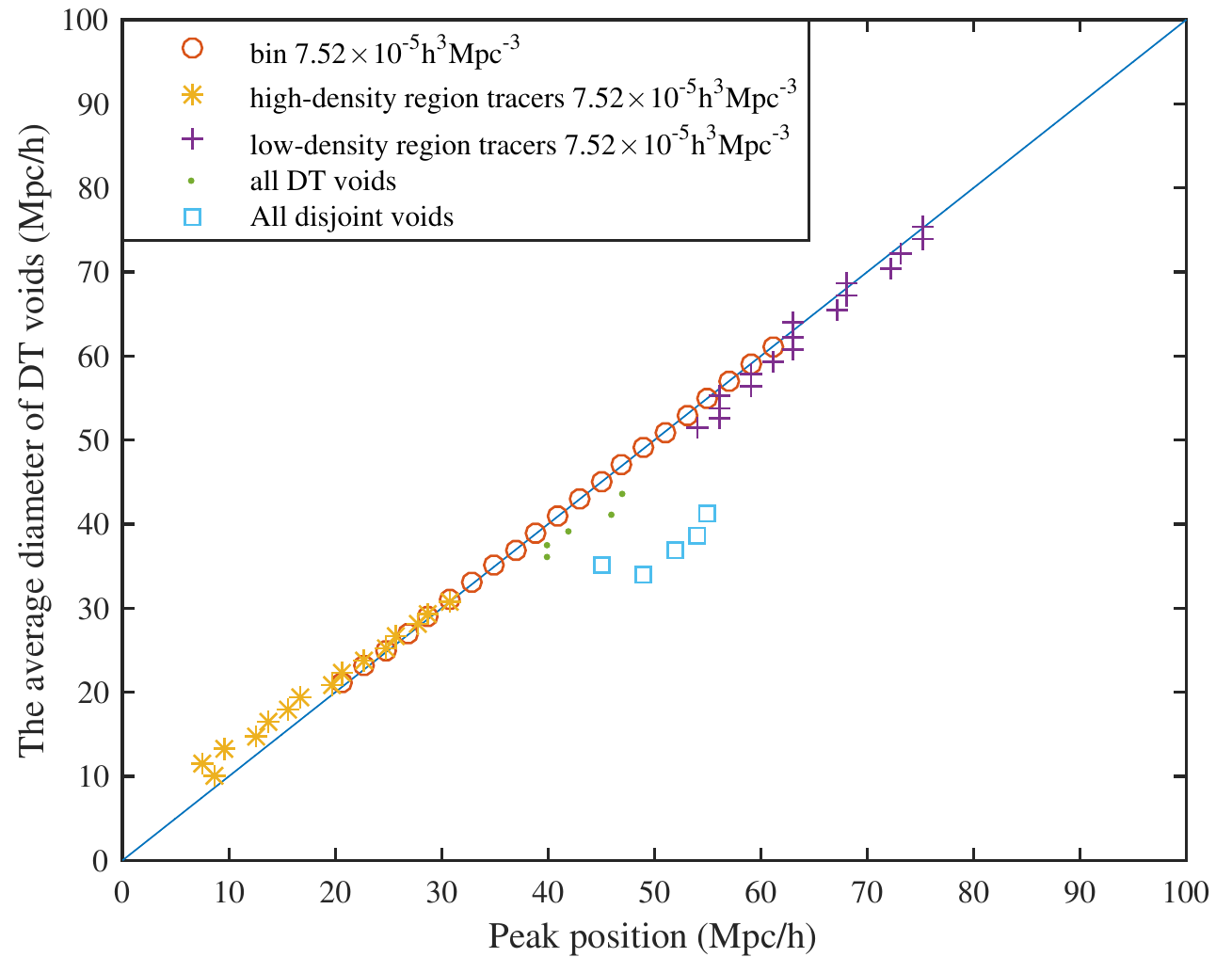}
  \caption{The average diameters of different data sample sets as functions of the peak positions of their two-point correlation functions. As shown, there is a strong correlation between the average diameter of a certain data sample set and the peak position of the two-point correlation function of this data sample set. In addition, it also show the relationships of the five disjoint void catalogues with different number density, which are marked by squares under the straight line due to the non-overlapping pattern.}
 \label{fig:peakposition}
\end{figure}

In Fig.\ref{fig:peakposition}, we plot the average diameters of the samples from different DT void populations and disjoint void populations as functions of the peak positions of their two-point correlation functions. It clearly shows that the peak position of two-point correlation function definitely corresponds to the average diameter of the samples for the overlapping DT voids. Whereas, for the disjoint voids, the positions of the peaks are larger than the average diameters of the sample sets, which is not difficult to understand. This is due to reason of non-overlapping.

\section{Discussion on the difference of the BAO detections via different tracers (mock galaxies, low/high-density region tracers)}
\label{sec:BAO_tracers difference}

\begin{figure}
 \includegraphics[width=0.9\linewidth]{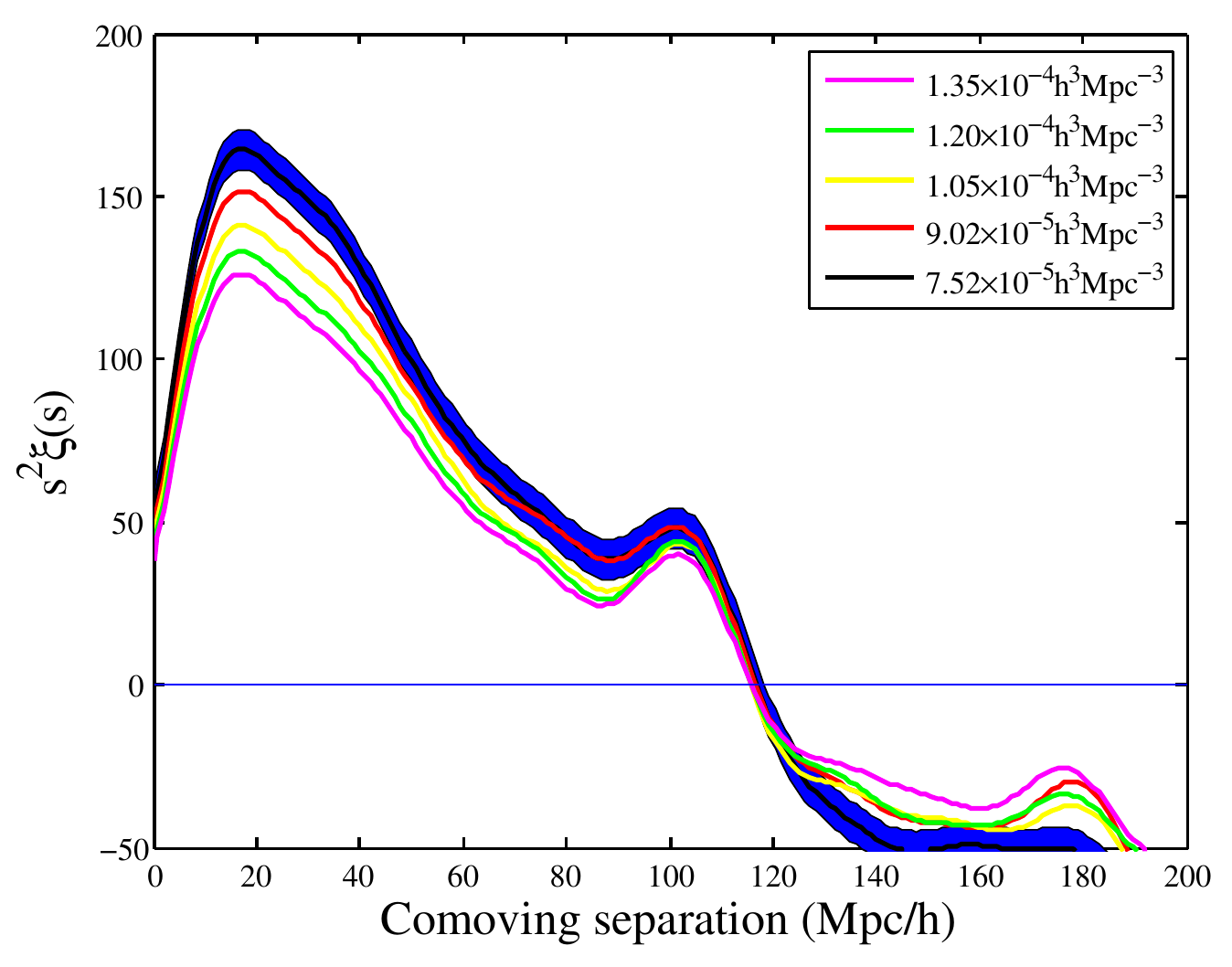}
  \includegraphics[width=0.9\linewidth]{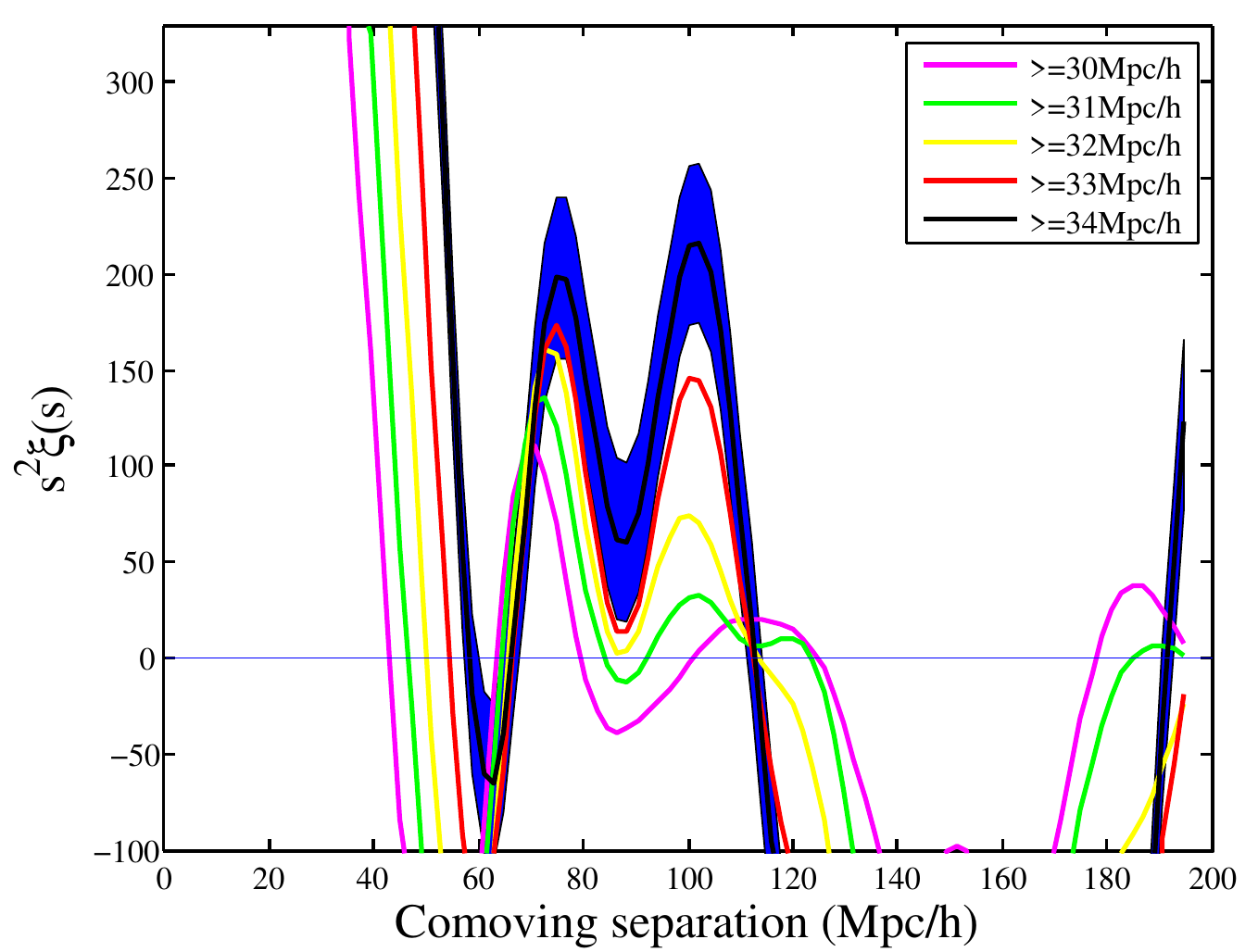}
  \flushleft
  \vspace{-0.2cm}\hspace{-0.15cm}\includegraphics[width=0.91\linewidth]{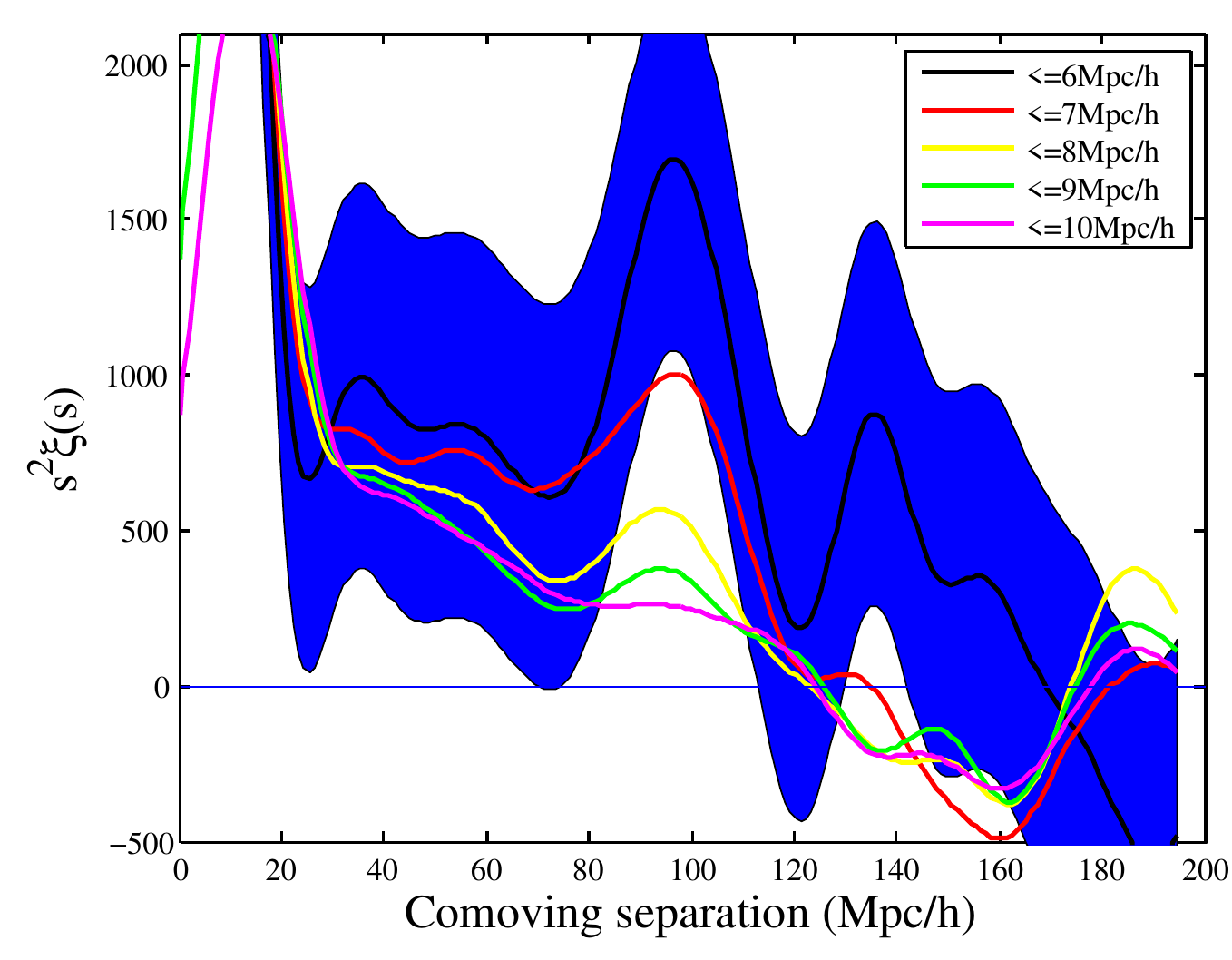}
  \caption{The two-point correlation functions of mock galaxies (the top panel) \& low/high-density region tracers (the middle/bottom panel). the curves are a part of the two-point correlation functions in Fig.[\ref{fig:halo}, \ref{fig:low_density_tracers}, \ref{fig:BAO}], fitted using the Gaussian Process Regression (GPR) method. The shaded regions show the $2\sigma$ errors from the Gaussian Process Regression method.}
  \label{fig:BAO_tracers}
\end{figure}


Our work rely on such an unprecedented large-scale high-resolution $N$-body simulation. Due to the huge cost, it is unrealistic to run numerous such $N$-body simulations with different seed initial conditions, and then by the procedure introduced in section \ref{sec:Mock galaxy} to construct plenty of mock galaxy catalogues with a specified number density, and finally by calculating the average values and variances of these independent data to eliminate systematic errors and give the error bars (similar to the method of \citealt{2016MNRAS.459.4020L}). However, we can give a proposal to overcome this weakness to get multiple mock galaxy catalogues with a specified number density, by the following two steps:

\begin{itemize}
    \item Firstly, according to the method described in the section \ref{sec:Mock galaxy}, from the original halo catalogue we select out the massive halos with a certain amount, larger than the sample amounts of the final mock galaxy catalogues with the specified number density, as the pretreated mock galaxy catalogue.

    \item Secondly, after the pretreatment, we then randomly select out the samples with the specified amount by sampling without replacement method
    to construct a mock galaxy catalogue with the specified number density. And by repeating this step we can get more mock galaxy catalogues with the specified number density.
\end{itemize}

Of course there are something worth noting. e.g. in order to avoid getting too similar or the same mock galaxy catalogues, we need to ensure the pretreated mock galaxy catalogue capacity being large enough; and we'd better give massive halos with greater likelihood of being selected and so on. Again, for the same reason that we are more concerned about BAO detections from different tracers, more detailed exploration is reserved for future work.

For a better analysis of the difference, part of the two-point correlation functions in Fig.[\ref{fig:halo}, \ref{fig:low_density_tracers}, \ref{fig:BAO}], in the best case for BAO signal detection, are fitted by a model-independent fitting method, i.e. the Gaussian Process Regression (GPR) method (cf. Fig.\ref{fig:BAO_tracers}). And then we can identify some important features. In the bottom panel of Fig.\ref{fig:BAO_tracers}, for the high-density region tracers, we can witness the process that as the radius cut decreases, the BAO signal arises and grows, whereas, for the low-density region tracers, we can witness a similar but actually the opposite process that as the radius cut increases to the optimal cut, the BAO signal begins to arises and grows to the strongest (cf. middle panel of Fig.\ref{fig:BAO_tracers}). Surprisingly, in the best case, the intensity of the BAO signals (the heights of the BAO peaks), detected respectively by mock galaxies, low-density region tracers, and high-density region tracers, almost in sequence increase by an order of magnitude (cf. Fig.\ref{fig:BAO_tracers}).

The process of baryon acoustic oscillations in the early Universe leads to the material density ripples, the structure of which is mainly configured on the regions with high material density, i.e. the regions with high galaxy number density, in the Universe. As shown in former sections, the spatial distribution of high-density region tracers (small DT voids), can reliably characterize the contour structure of the high-density regions in the Universe (cf. \ref{fig:tracers}). Therefore, one plausible explanation of our results is that the BAO structure is mainly configured on these high-density regions and we can extract the high-dense structure by the locations of high-density region tracers (small DT voids), as a consequence, the BAO intensity (BAO peak) shown on the two-point correlation function of the high-density tracers can be greatly enhanced. Indeed, the SDSS Luminous Red Galaxy (LRG) samples utilized in \citealt{2005ApJ...633..560E}, when the BAO feature was for the first time detected, are distributed in the high-density regions (cf. \citealt{2015MNRAS.450.1836K}).

The material ripples of BAO overlap with each other so that the distribution of low-density region tracers, which are located in the regions spatially complementary to the high-density regions (cf. \ref{fig:tracers}), can also be configured with the BAO distribution information at BAO scales. As a result, we can also detect significant BAO peak on the two-point correlation function of the low-density region tracers. However, this requires a reasonable tuning of a threshold parameter of the samples. Here, we tune the DT void radius cut (the minimum radius of large DT void samples) to the optimal value, such that the BAO signal emerges and becomes most significant. Surprisingly, the optimal BAO signal shown on the two-point correlation of the low-density region tracers is also more prominent than the signal detected by mock galaxies (massive halos) (cf. \ref{fig:BAO_tracers}).

\section{conclusion and discussion}
\label{sec:SUMMARY}
As the distinctive features of the cosmic web, the cosmic voids are the large and low-density regions which contain less galaxies/halos and are surrounded by filaments, walls and relatively denser knots of galaxies/halos. Due to the ambiguity of void definition, there is not an unequivocal method for systematically searching cosmic voids in both surveys and simulations, so that voids can be defined in many different ways for different purposes.

In this work, relying on a large-scale high-precision $N$-body simulation and by a halo finding procedure using spherical overdensity approach we find 27.758 million dark matter halos. We propose a HOD model to construct five mock galaxy catalogues with different number density. And we take advantage of a Delaunay Triangulation void finder (\texttt{DIVE}) to construct a large number of empty overlapping circumspheres (named as DT voids) constrained by tetrahedra of the mock galaxies/halos. By this void definition, the number of DT voids is about 7 times the mock galaxy population, which permits us to perform reliable statistical studies, in particular clustering analysis, based on these DT voids.

As a matter of fact, the total volume of the DT voids exceeds 200 times the volume of the simulation box in the context of this work. Although the DT voids are overlapping seriously, we can still straightforwardly construct another type of voids, the non-overlapping spheres called disjoint voids, based on these overlapping DT voids.

We conduct some fundamental statistical studies and clustering analysis on these DT voids and disjoint voids. We show that these statistics of the DT voids and disjoint voids are strongly correlated with the number density of the corresponding mock galaxies. But it is interesting to note that although the space occupied by the all disjoint voids reduces corresponding to the decline in the number density of the mock galaxies, this effect is very slight with the total volume of these disjoint voids maintained at about $45\%$ of the simulation box volume. And we also show that the peak position of the DT void two-point correlation function has a very strong correlation with the average diameter of the void samples.

Even if DT voids are very overlapping, the sizes of their radii can be very good to reflect the halo number density of their interior and surrounding areas, which permits us to give a reliable definition of different density region tracers. We show that the locations of large DT voids well trace the low-density regions and the locations of small DT voids well trace the high-density regions.

Moreover, we discuss in detail the features of the two-point correlation functions of different DT void populations. And, we further investigate the BAO detections by the two-point correlation functions of low/high-density region tracers, corresponding to the mock galaxy number density of $7.52\times10^{-5}$ $h^3$ Mpc $^{-3}$. We also confirm that for low-density region tracers there is a optimal void radius cut making the BAO signal the strongest. Our results show that for the mock galaxy catalogue with the number density of $7.52\times10^{-5}$ $h^3$ Mpc$^{-3}$ the optimal void radius cut is $\sim34$ Mpc$/h$. However, for the high-density region tracers, the different scenario is shown. We find that as the void radius cut decreases the BAO signal arises and grows continuously, but at the same time with the decline in the number of the data samples the noise becomes more and more prominent.

In order to better compare the BAO signal detections by mock galaxies and low/high-density region tracers, we use the Gaussian Process Regression (GPR) method to fit the two-point correlation functions of the five mock galaxy catalogues, and, in the optimal-case scenarios for the BAO signals, to fit part of the two-point correlation functions of the low/high-density region tracers (constructed by the mock galaxy catalogue with number density of $7.52\times10^{-5}$ $h^3$ Mpc$^{-3}$) respectively. Our results show that the BAO intensities (the heights of the BAO peaks) obtained by the three different ways in sequence increase by an order of magnitude.

We have two sets of $N$-body simulation data (one from TianZero and another from TianNu), which in principle permits us to study the effects of neutrinos on the large-scale structure formation of the Universe. Based on the two sets of data some pioneering works have been done, e.g. the effect of cosmic neutrinos on halo mass in \citealt{2017NatAs...1E.143Y}; measurement of the cold dark matter-neutrino dipole in \citealt{2016arXiv161009354I}.

We note that the DT voids, further constrained by imposing the circumspheres to be empty based on the tetrahedra of galaxies, are expected to encode higher order statistics (see \citealt{2016PhRvL.116q1301K}; \citealt{2016MNRAS.459.2670Z}; \citealt{2016MNRAS.459.4020L}), which strongly depends on gravitational evolution of the morphology of the cosmic web and provides an effective and powerful tool to study the large-scale structure formation history of the Universe. By analyzing the clustering of the DT voids and investigating their statistical properties, we expect to extract the crucial information of the neutrino effects on the large-scale structure formation of the Universe more efficiently and we will apply this technique for further studies of the neutrinos effects in cosmology.

\section*{Acknowledgements}
 Y.L. would like to thank Ue-Li Pen for helpful discussions and several useful suggestions. This work is supported by the National Science Foundation of China (Grants No. 11528306,11573006), the Fundamental ResearchFunds for the Central Universities, the Special Program for Applied Research on Super Computation of the NSFC-Guangdong Joint Fund (the second phase), and National Key R\&D Program of China (2017YFA0402600).


\bibliographystyle{aasjournal}  
\bibliography{ref}

\begin{thebibliography}{}
\expandafter\ifx\csname natexlab\endcsname\relax\def\natexlab#1{#1}\fi

\bibitem[{{Achitouv}(2016)}]{2016PhRvD..94j3524A}
{Achitouv}, I. 2016, \prd, 94, 103524

\bibitem[{{Alam} {et~al.}(2017){Alam}, {Ata}, {Bailey}, {Beutler}, {Bizyaev},
  {Blazek}, {Bolton}, {Brownstein}, {Burden}, {Chuang}, {Comparat}, {Cuesta},
  {Dawson}, {Eisenstein}, {Escoffier}, {Gil-Mar{\'{\i}}n}, {Grieb}, {Hand},
  {Ho}, {Kinemuchi}, {Kirkby}, {Kitaura}, {Malanushenko}, {Malanushenko},
  {Maraston}, {McBride}, {Nichol}, {Olmstead}, {Oravetz}, {Padmanabhan},
  {Palanque-Delabrouille}, {Pan}, {Pellejero-Ibanez}, {Percival}, {Petitjean},
  {Prada}, {Price-Whelan}, {Reid}, {Rodr{\'{\i}}guez-Torres}, {Roe}, {Ross},
  {Ross}, {Rossi}, {Rubi{\~n}o-Mart{\'{\i}}n}, {Saito}, {Salazar-Albornoz},
  {Samushia}, {S{\'a}nchez}, {Satpathy}, {Schlegel}, {Schneider},
  {Sc{\'o}ccola}, {Seo}, {Sheldon}, {Simmons}, {Slosar}, {Strauss}, {Swanson},
  {Thomas}, {Tinker}, {Tojeiro}, {Maga{\~n}a}, {Vazquez}, {Verde}, {Wake},
  {Wang}, {Weinberg}, {White}, {Wood-Vasey}, {Y{\`e}che}, {Zehavi}, {Zhai}, \&
  {Zhao}}]{2017MNRAS.470.2617A}
{Alam}, S., {Ata}, M., {Bailey}, S., {et~al.} 2017, \mnras, 470, 2617

\bibitem[{{Albrecht} {et~al.}(2006){Albrecht}, {Bernstein}, {Cahn}, {Freedman},
  {Hewitt}, {Hu}, {Huth}, {Kamionkowski}, {Kolb}, {Knox}, {Mather}, {Staggs},
  \& {Suntzeff}}]{2006astro.ph..9591A}
{Albrecht}, A., {Bernstein}, G., {Cahn}, R., {et~al.} 2006, ArXiv Astrophysics
  e-prints, astro-ph/0609591

\bibitem[{{Alcock} \& {Paczynski}(1979)}]{1979Natur.281..358A}
{Alcock}, C., \& {Paczynski}, B. 1979, \nat, 281, 358

\bibitem[{{Anderson} {et~al.}(2014{\natexlab{a}}){Anderson}, {Aubourg},
  {Bailey}, {Beutler}, {Bhardwaj}, {Blanton}, {Bolton}, {Brinkmann},
  {Brownstein}, {Burden}, {Chuang}, {Cuesta}, {Dawson}, {Eisenstein},
  {Escoffier}, {Gunn}, {Guo}, {Ho}, {Honscheid}, {Howlett}, {Kirkby}, {Lupton},
  {Manera}, {Maraston}, {McBride}, {Mena}, {Montesano}, {Nichol}, {Nuza},
  {Olmstead}, {Padmanabhan}, {Palanque-Delabrouille}, {Parejko}, {Percival},
  {Petitjean}, {Prada}, {Price-Whelan}, {Reid}, {Roe}, {Ross}, {Ross}, {Sabiu},
  {Saito}, {Samushia}, {S{\'a}nchez}, {Schlegel}, {Schneider}, {Scoccola},
  {Seo}, {Skibba}, {Strauss}, {Swanson}, {Thomas}, {Tinker}, {Tojeiro},
  {Maga{\~n}a}, {Verde}, {Wake}, {Weaver}, {Weinberg}, {White}, {Xu},
  {Y{\`e}che}, {Zehavi}, \& {Zhao}}]{2014MNRAS.441...24A}
{Anderson}, L., {Aubourg}, {\'E}., {Bailey}, S., {et~al.} 2014{\natexlab{a}},
  \mnras, 441, 24

\bibitem[{{Anderson} {et~al.}(2014{\natexlab{b}}){Anderson}, {Aubourg},
  {Bailey}, {Beutler}, {Bolton}, {Brinkmann}, {Brownstein}, {Chuang}, {Cuesta},
  {Dawson}, {Eisenstein}, {Ho}, {Honscheid}, {Kazin}, {Kirkby}, {Manera},
  {McBride}, {Mena}, {Nichol}, {Olmstead}, {Padmanabhan},
  {Palanque-Delabrouille}, {Percival}, {Prada}, {Ross}, {Ross}, {S{\'a}nchez},
  {Samushia}, {Schlegel}, {Schneider}, {Seo}, {Strauss}, {Thomas}, {Tinker},
  {Tojeiro}, {Verde}, {Wake}, {Weinberg}, {Xu}, \&
  {Yeche}}]{2014MNRAS.439...83A}
{Anderson}, L., {Aubourg}, E., {Bailey}, S., {et~al.} 2014{\natexlab{b}},
  \mnras, 439, 83

\bibitem[{{Ata} {et~al.}(2017){Ata}, {Baumgarten}, {Bautista}, {Beutler},
  {Bizyaev}, {Blanton}, {Blazek}, {Bolton}, {Brinkmann}, {Brownstein},
  {Burtin}, {Chuang}, {Comparat}, {Dawson}, {de la Macorra}, {Du}, {du Mas des
  Bourboux}, {Eisenstein}, {Gil-Marin}, {Grabowski}, {Guy}, {Hand}, {Ho},
  {Hutchinson}, {Ivanov}, {Kitaura}, {Kneib}, {Laurent}, {Le Goff}, {McEwen},
  {Mueller}, {Myers}, {Newman}, {Palanque-Delabrouille}, {Pan}, {Paris},
  {Pellejero-Ibanez}, {Percival}, {Petitjean}, {Prada}, {Prakash},
  {Rodriguez-Torres}, {Ross}, {Rossi}, {Ruggeri}, {Sanchez}, {Satpathy},
  {Schlegel}, {Schneider}, {Seo}, {Slosar}, {Streblyanska}, {Tinker},
  {Tojeiro}, {Vargas Magana}, {Vivek}, {Wang}, {Yeche}, {Yu}, {Zarrouk},
  {Zhao}, {Zhao}, \& {Zhu}}]{2017arXiv170506373A}
{Ata}, M., {Baumgarten}, F., {Bautista}, J., {et~al.} 2017, ArXiv e-prints,
  arXiv:1705.06373

\bibitem[{{Barreira} {et~al.}(2015){Barreira}, {Cautun}, {Li}, {Baugh}, \&
  {Pascoli}}]{2015JCAP...08..028B}
{Barreira}, A., {Cautun}, M., {Li}, B., {Baugh}, C.~M., \& {Pascoli}, S. 2015,
  \jcap, 8, 028

\bibitem[{{Bautista} {et~al.}(2017){Bautista}, {Busca}, {Guy}, {Rich},
  {Blomqvist}, {du Mas des Bourboux}, {Pieri}, {Font-Ribera}, {Bailey},
  {Delubac}, {Kirkby}, {Le Goff}, {Margala}, {Slosar}, {Vazquez}, {Brownstein},
  {Dawson}, {Eisenstein}, {Miralda-Escud{\'e}}, {Noterdaeme},
  {Palanque-Delabrouille}, {P{\^a}ris}, {Petitjean}, {Ross}, {Schneider},
  {Weinberg}, \& {Y{\`e}che}}]{2017A&A...603A..12B}
{Bautista}, J.~E., {Busca}, N.~G., {Guy}, J., {et~al.} 2017, \aap, 603, A12

\bibitem[{{Benitez} {et~al.}(2014){Benitez}, {Dupke}, {Moles}, {Sodre},
  {Cenarro}, {Marin-Franch}, {Taylor}, {Cristobal}, {Fernandez-Soto}, {Mendes
  de Oliveira}, {Cepa-Nogue}, {Abramo}, {Alcaniz}, {Overzier},
  {Hernandez-Monteagudo}, {Alfaro}, {Kanaan}, {Carvano}, {Reis}, {Martinez
  Gonzalez}, {Ascaso}, {Ballesteros}, {Xavier}, {Varela}, {Ederoclite},
  {Vazquez Ramio}, {Broadhurst}, {Cypriano}, {Angulo}, {Diego}, {Zandivarez},
  {Diaz}, {Melchior}, {Umetsu}, {Spinelli}, {Zitrin}, {Coe}, {Yepes}, {Vielva},
  {Sahni}, {Marcos-Caballero}, {Shu Kitaura}, {Maroto}, {Masip}, {Tsujikawa},
  {Carneiro}, {Gonzalez Nuevo}, {Carvalho}, {Reboucas}, {Carvalho}, {Abdalla},
  {Bernui}, {Pigozzo}, {Ferreira}, {Chandrachani Devi}, {Bengaly}, {Campista},
  {Amorim}, {Asari}, {Bongiovanni}, {Bonoli}, {Bruzual}, {Cardiel}, {Cava},
  {Cid Fernandes}, {Coelho}, {Cortesi}, {Delgado}, {Diaz Garcia}, {Espinosa},
  {Galliano}, {Gonzalez-Serrano}, {Falcon-Barroso}, {Fritz}, {Fernandes},
  {Gorgas}, {Hoyos}, {Jimenez-Teja}, {Lopez-Aguerri}, {Lopez-San Juan},
  {Mateus}, {Molino}, {Novais}, {OMill}, {Oteo}, {Perez-Gonzalez}, {Poggianti},
  {Proctor}, {Ricciardelli}, {Sanchez-Blazquez}, {Storchi-Bergmann}, {Telles},
  {Schoennell}, {Trujillo}, {Vazdekis}, {Viironen}, {Daflon},
  {Aparicio-Villegas}, {Rocha}, {Ribeiro}, {Borges}, {Martins}, {Marcolino},
  {Martinez-Delgado}, {Perez-Torres}, {Siffert}, {Calvao}, {Sako}, {Kessler},
  {Alvarez-Candal}, {De Pra}, {Roig}, {Lazzaro}, {Gorosabel}, {Lopes de
  Oliveira}, {Lima-Neto}, {Irwin}, {Liu}, {Alvarez}, {Balmes}, {Chueca},
  {Costa-Duarte}, {da Costa}, {Dantas}, {Diaz}, {Fabregat}, {Ferrari},
  {Gavela}, {Gracia}, {Gruel}, {Gutierrez}, {Guzman}, {Hernandez-Fernandez},
  {Herranz}, {Hurtado-Gil}, {Jablonsky}, {Laporte}, {Le Tiran}, {Licandro},
  {Lima}, {Martin}, {Martinez}, {Montero}, {Penteado}, {Pereira}, {Peris},
  {Quilis}, {Sanchez-Portal}, {Soja}, {Solano}, {Torra}, \&
  {Valdivielso}}]{2014arXiv1403.5237B}
{Benitez}, N., {Dupke}, R., {Moles}, M., {et~al.} 2014, ArXiv e-prints,
  arXiv:1403.5237

\bibitem[{{Bennett} {et~al.}(2003){Bennett}, {Halpern}, {Hinshaw}, {Jarosik},
  {Kogut}, {Limon}, {Meyer}, {Page}, {Spergel}, {Tucker}, {Wollack}, {Wright},
  {Barnes}, {Greason}, {Hill}, {Komatsu}, {Nolta}, {Odegard}, {Peiris},
  {Verde}, \& {Weiland}}]{2003ApJS..148....1B}
{Bennett}, C.~L., {Halpern}, M., {Hinshaw}, G., {et~al.} 2003, \apjs, 148, 1

\bibitem[{{Betancort-Rijo}(1990)}]{1990MNRAS.246..608B}
{Betancort-Rijo}, J. 1990, \mnras, 246, 608

\bibitem[{{Betancort-Rijo} \&
  {L{\'o}pez-Corredoira}(2002)}]{2002ApJ...566..623B}
{Betancort-Rijo}, J., \& {L{\'o}pez-Corredoira}, M. 2002, \apj, 566, 623

\bibitem[{{Betancort-Rijo} {et~al.}(2009){Betancort-Rijo}, {Patiri}, {Prada},
  \& {Romano}}]{2009MNRAS.400.1835B}
{Betancort-Rijo}, J., {Patiri}, S.~G., {Prada}, F., \& {Romano}, A.~E. 2009,
  \mnras, 400, 1835

\bibitem[{{Beutler} {et~al.}(2011){Beutler}, {Blake}, {Colless}, {Jones},
  {Staveley-Smith}, {Campbell}, {Parker}, {Saunders}, \&
  {Watson}}]{2011MNRAS.416.3017B}
{Beutler}, F., {Blake}, C., {Colless}, M., {et~al.} 2011, \mnras, 416, 3017

\bibitem[{{Blake} \& {Glazebrook}(2003)}]{2003ApJ...594..665B}
{Blake}, C., \& {Glazebrook}, K. 2003, \apj, 594, 665

\bibitem[{{Bond} \& {Efstathiou}(1984)}]{1984ApJ...285L..45B}
{Bond}, J.~R., \& {Efstathiou}, G. 1984, \apjl, 285, L45

\bibitem[{{Bos} {et~al.}(2012){Bos}, {van de Weygaert}, {Dolag}, \&
  {Pettorino}}]{2012MNRAS.426..440B}
{Bos}, E.~G.~P., {van de Weygaert}, R., {Dolag}, K., \& {Pettorino}, V. 2012,
  \mnras, 426, 440

\bibitem[{{Brunino} {et~al.}(2007){Brunino}, {Trujillo}, {Pearce}, \&
  {Thomas}}]{2007MNRAS.375..184B}
{Brunino}, R., {Trujillo}, I., {Pearce}, F.~R., \& {Thomas}, P.~A. 2007,
  \mnras, 375, 184

\bibitem[{{Busca} {et~al.}(2013){Busca}, {Delubac}, {Rich}, {Bailey},
  {Font-Ribera}, {Kirkby}, {Le Goff}, {Pieri}, {Slosar}, {Aubourg}, {Bautista},
  {Bizyaev}, {Blomqvist}, {Bolton}, {Bovy}, {Brewington}, {Borde}, {Brinkmann},
  {Carithers}, {Croft}, {Dawson}, {Ebelke}, {Eisenstein}, {Hamilton}, {Ho},
  {Hogg}, {Honscheid}, {Lee}, {Lundgren}, {Malanushenko}, {Malanushenko},
  {Margala}, {Maraston}, {Mehta}, {Miralda-Escud{\'e}}, {Myers}, {Nichol},
  {Noterdaeme}, {Olmstead}, {Oravetz}, {Palanque-Delabrouille}, {Pan},
  {P{\^a}ris}, {Percival}, {Petitjean}, {Roe}, {Rollinde}, {Ross}, {Rossi},
  {Schlegel}, {Schneider}, {Shelden}, {Sheldon}, {Simmons}, {Snedden},
  {Tinker}, {Viel}, {Weaver}, {Weinberg}, {White}, {Y{\`e}che}, \&
  {York}}]{2013A&A...552A..96B}
{Busca}, N.~G., {Delubac}, T., {Rich}, J., {et~al.} 2013, \aap, 552, A96

\bibitem[{{Cai} {et~al.}(2014){Cai}, {Li}, {Cole}, {Frenk}, \&
  {Neyrinck}}]{2014MNRAS.439.2978C}
{Cai}, Y.-C., {Li}, B., {Cole}, S., {Frenk}, C.~S., \& {Neyrinck}, M. 2014,
  \mnras, 439, 2978

\bibitem[{{Cai} {et~al.}(2015){Cai}, {Padilla}, \& {Li}}]{2015MNRAS.451.1036C}
{Cai}, Y.-C., {Padilla}, N., \& {Li}, B. 2015, \mnras, 451, 1036

\bibitem[{{Cautun} \& {van de Weygaert}(2011)}]{2011ascl.soft05003C}
{Cautun}, M.~C., \& {van de Weygaert}, R. 2011, {The DTFE public software: The
  Delaunay Tessellation Field Estimator code}, Astrophysics Source Code
  Library, , , arXiv:1105.0370

\bibitem[{{Chen}(2015)}]{2015IAUGA..2252187C}
{Chen}, X. 2015, IAU General Assembly, 22, 2252187

\bibitem[{{Clampitt} {et~al.}(2013){Clampitt}, {Cai}, \&
  {Li}}]{2013MNRAS.431..749C}
{Clampitt}, J., {Cai}, Y.-C., \& {Li}, B. 2013, \mnras, 431, 749

\bibitem[{{Clampitt} \& {Jain}(2015)}]{2015MNRAS.454.3357C}
{Clampitt}, J., \& {Jain}, B. 2015, \mnras, 454, 3357

\bibitem[{{Clampitt} {et~al.}(2016){Clampitt}, {Jain}, \&
  {S{\'a}nchez}}]{2016MNRAS.456.4425C}
{Clampitt}, J., {Jain}, B., \& {S{\'a}nchez}, C. 2016, \mnras, 456, 4425

\bibitem[{{Cole} {et~al.}(2005){Cole}, {Percival}, {Peacock}, {Norberg},
  {Baugh}, {Frenk}, {Baldry}, {Bland-Hawthorn}, {Bridges}, {Cannon}, {Colless},
  {Collins}, {Couch}, {Cross}, {Dalton}, {Eke}, {De Propris}, {Driver},
  {Efstathiou}, {Ellis}, {Glazebrook}, {Jackson}, {Jenkins}, {Lahav}, {Lewis},
  {Lumsden}, {Maddox}, {Madgwick}, {Peterson}, {Sutherland}, \&
  {Taylor}}]{2005MNRAS.362..505C}
{Cole}, S., {Percival}, W.~J., {Peacock}, J.~A., {et~al.} 2005, \mnras, 362,
  505

\bibitem[{{Colless} {et~al.}(2001){Colless}, {Dalton}, {Maddox}, {Sutherland},
  {Norberg}, {Cole}, {Bland-Hawthorn}, {Bridges}, {Cannon}, {Collins}, {Couch},
  {Cross}, {Deeley}, {De Propris}, {Driver}, {Efstathiou}, {Ellis}, {Frenk},
  {Glazebrook}, {Jackson}, {Lahav}, {Lewis}, {Lumsden}, {Madgwick}, {Peacock},
  {Peterson}, {Price}, {Seaborne}, \& {Taylor}}]{2001MNRAS.328.1039C}
{Colless}, M., {Dalton}, G., {Maddox}, S., {et~al.} 2001, \mnras, 328, 1039

\bibitem[{{Croton} {et~al.}(2005){Croton}, {Farrar}, {Norberg}, {Colless},
  {Peacock}, {Baldry}, {Baugh}, {Bland-Hawthorn}, {Bridges}, {Cannon}, {Cole},
  {Collins}, {Couch}, {Dalton}, {De Propris}, {Driver}, {Efstathiou}, {Ellis},
  {Frenk}, {Glazebrook}, {Jackson}, {Lahav}, {Lewis}, {Lumsden}, {Maddox},
  {Madgwick}, {Peterson}, {Sutherland}, \& {Taylor}}]{2005MNRAS.356.1155C}
{Croton}, D.~J., {Farrar}, G.~R., {Norberg}, P., {et~al.} 2005, \mnras, 356,
  1155

\bibitem[{{Davis} \& {Peebles}(1983)}]{1983ApJ...267..465D}
{Davis}, M., \& {Peebles}, P.~J.~E. 1983, \apj, 267, 465

\bibitem[{{de Jong} {et~al.}(2012){de Jong}, {Bellido-Tirado}, {Chiappini},
  {Depagne}, {Haynes}, {Johl}, {Schnurr}, {Schwope}, {Walcher}, {Dionies},
  {Haynes}, {Kelz}, {Kitaura}, {Lamer}, {Minchev}, {M{\"u}ller}, {Nuza},
  {Olaya}, {Piffl}, {Popow}, {Steinmetz}, {Ural}, {Williams}, {Winkler},
  {Wisotzki}, {Ansorge}, {Banerji}, {Gonzalez Solares}, {Irwin}, {Kennicutt},
  {King}, {McMahon}, {Koposov}, {Parry}, {Sun}, {Walton}, {Finger}, {Iwert},
  {Krumpe}, {Lizon}, {Vincenzo}, {Amans}, {Bonifacio}, {Cohen}, {Francois},
  {Jagourel}, {Mignot}, {Royer}, {Sartoretti}, {Bender}, {Grupp}, {Hess},
  {Lang-Bardl}, {Muschielok}, {B{\"o}hringer}, {Boller}, {Bongiorno}, {Brusa},
  {Dwelly}, {Merloni}, {Nandra}, {Salvato}, {Pragt}, {Navarro}, {Gerlofsma},
  {Roelfsema}, {Dalton}, {Middleton}, {Tosh}, {Boeche}, {Caffau}, {Christlieb},
  {Grebel}, {Hansen}, {Koch}, {Ludwig}, {Quirrenbach}, {Sbordone}, {Seifert},
  {Thimm}, {Trifonov}, {Helmi}, {Trager}, {Feltzing}, {Korn}, \&
  {Boland}}]{2012SPIE.8446E..0TD}
{de Jong}, R.~S., {Bellido-Tirado}, O., {Chiappini}, C., {et~al.} 2012, in
  \procspie, Vol. 8446, Ground-based and Airborne Instrumentation for Astronomy
  IV, 84460T

\bibitem[{Delaunay(1934)}]{delaunay1934sphere}
Delaunay, B. 1934, Izv. Akad. Nauk SSSR, Otdelenie Matematicheskii i
  Estestvennyka Nauk, 7, 1

\bibitem[{{Delubac} {et~al.}(2015){Delubac}, {Bautista}, {Busca}, {Rich},
  {Kirkby}, {Bailey}, {Font-Ribera}, {Slosar}, {Lee}, {Pieri}, {Hamilton},
  {Aubourg}, {Blomqvist}, {Bovy}, {Brinkmann}, {Carithers}, {Dawson},
  {Eisenstein}, {Gontcho}, {Kneib}, {Le Goff}, {Margala}, {Miralda-Escud{\'e}},
  {Myers}, {Nichol}, {Noterdaeme}, {O'Connell}, {Olmstead},
  {Palanque-Delabrouille}, {P{\^a}ris}, {Petitjean}, {Ross}, {Rossi},
  {Schlegel}, {Schneider}, {Weinberg}, {Y{\`e}che}, \&
  {York}}]{2015A&A...574A..59D}
{Delubac}, T., {Bautista}, J.~E., {Busca}, N.~G., {et~al.} 2015, \aap, 574, A59

\bibitem[{{Drinkwater} {et~al.}(2010){Drinkwater}, {Jurek}, {Blake}, {Woods},
  {Pimbblet}, {Glazebrook}, {Sharp}, {Pracy}, {Brough}, {Colless}, {Couch},
  {Croom}, {Davis}, {Forbes}, {Forster}, {Gilbank}, {Gladders}, {Jelliffe},
  {Jones}, {Li}, {Madore}, {Martin}, {Poole}, {Small}, {Wisnioski}, {Wyder}, \&
  {Yee}}]{2010MNRAS.401.1429D}
{Drinkwater}, M.~J., {Jurek}, R.~J., {Blake}, C., {et~al.} 2010, \mnras, 401,
  1429

\bibitem[{{Einasto} {et~al.}(1991){Einasto}, {Einasto}, {Gramann}, \&
  {Saar}}]{1991MNRAS.248..593E}
{Einasto}, J., {Einasto}, M., {Gramann}, M., \& {Saar}, E. 1991, \mnras, 248,
  593

\bibitem[{{Eisenstein}(2005)}]{2005NewAR..49..360E}
{Eisenstein}, D.~J. 2005, \nar, 49, 360

\bibitem[{{Eisenstein} \& {Hu}(1998)}]{1998ApJ...496..605E}
{Eisenstein}, D.~J., \& {Hu}, W. 1998, \apj, 496, 605

\bibitem[{{Eisenstein} {et~al.}(2005){Eisenstein}, {Zehavi}, {Hogg},
  {Scoccimarro}, {Blanton}, {Nichol}, {Scranton}, {Seo}, {Tegmark}, {Zheng},
  {Anderson}, {Annis}, {Bahcall}, {Brinkmann}, {Burles}, {Castander},
  {Connolly}, {Csabai}, {Doi}, {Fukugita}, {Frieman}, {Glazebrook}, {Gunn},
  {Hendry}, {Hennessy}, {Ivezi{\'c}}, {Kent}, {Knapp}, {Lin}, {Loh}, {Lupton},
  {Margon}, {McKay}, {Meiksin}, {Munn}, {Pope}, {Richmond}, {Schlegel},
  {Schneider}, {Shimasaku}, {Stoughton}, {Strauss}, {SubbaRao}, {Szalay},
  {Szapudi}, {Tucker}, {Yanny}, \& {York}}]{2005ApJ...633..560E}
{Eisenstein}, D.~J., {Zehavi}, I., {Hogg}, D.~W., {et~al.} 2005, \apj, 633, 560

\bibitem[{{Eisenstein} {et~al.}(2011){Eisenstein}, {Weinberg}, {Agol},
  {Aihara}, {Allende Prieto}, {Anderson}, {Arns}, {Aubourg}, {Bailey},
  {Balbinot}, \& et~al.}]{2011AJ....142...72E}
{Eisenstein}, D.~J., {Weinberg}, D.~H., {Agol}, E., {et~al.} 2011, \aj, 142, 72

\bibitem[{{Emberson} {et~al.}(2017){Emberson}, {Yu}, {Inman}, {Zhang}, {Pen},
  {Harnois-D{\'e}raps}, {Yuan}, {Teng}, {Zhu}, {Chen}, \&
  {Xing}}]{2017RAA....17...85E}
{Emberson}, J.~D., {Yu}, H.-R., {Inman}, D., {et~al.} 2017, Research in
  Astronomy and Astrophysics, 17, 085

\bibitem[{{Goldberg} \& {Vogeley}(2004)}]{2004ApJ...605....1G}
{Goldberg}, D.~M., \& {Vogeley}, M.~S. 2004, \apj, 605, 1

\bibitem[{{Goldwirth} {et~al.}(1995){Goldwirth}, {da Costa}, \& {van de
  Weygaert}}]{1995MNRAS.275.1185G}
{Goldwirth}, D.~S., {da Costa}, L.~N., \& {van de Weygaert}, R. 1995, \mnras,
  275, 1185

\bibitem[{{Granett} {et~al.}(2015){Granett}, {Kov{\'a}cs}, \&
  {Hawken}}]{2015MNRAS.454.2804G}
{Granett}, B.~R., {Kov{\'a}cs}, A., \& {Hawken}, A.~J. 2015, \mnras, 454, 2804

\bibitem[{{Granett} {et~al.}(2008){Granett}, {Neyrinck}, \&
  {Szapudi}}]{2008ApJ...683L..99G}
{Granett}, B.~R., {Neyrinck}, M.~C., \& {Szapudi}, I. 2008, \apjl, 683, L99

\bibitem[{{Hamaus} {et~al.}(2014{\natexlab{a}}){Hamaus}, {Sutter}, \&
  {Wandelt}}]{2014PhRvL.112y1302H}
{Hamaus}, N., {Sutter}, P.~M., \& {Wandelt}, B.~D. 2014{\natexlab{a}}, Physical
  Review Letters, 112, 251302

\bibitem[{{Hamaus} {et~al.}(2014{\natexlab{b}}){Hamaus}, {Wandelt}, {Sutter},
  {Lavaux}, \& {Warren}}]{2014PhRvL.112d1304H}
{Hamaus}, N., {Wandelt}, B.~D., {Sutter}, P.~M., {Lavaux}, G., \& {Warren},
  M.~S. 2014{\natexlab{b}}, Physical Review Letters, 112, 041304

\bibitem[{{Hamilton}(1993)}]{1993ApJ...417...19H}
{Hamilton}, A.~J.~S. 1993, \apj, 417, 19

\bibitem[{{Harnois-D{\'e}raps} {et~al.}(2013){Harnois-D{\'e}raps}, {Pen},
  {Iliev}, {Merz}, {Emberson}, \& {Desjacques}}]{2013MNRAS.436..540H}
{Harnois-D{\'e}raps}, J., {Pen}, U.-L., {Iliev}, I.~T., {et~al.} 2013, \mnras,
  436, 540

\bibitem[{{Hewett}(1982)}]{1982MNRAS.201..867H}
{Hewett}, P.~C. 1982, \mnras, 201, 867

\bibitem[{{Hinshaw} {et~al.}(2013){Hinshaw}, {Larson}, {Komatsu}, {Spergel},
  {Bennett}, {Dunkley}, {Nolta}, {Halpern}, {Hill}, {Odegard}, {Page}, {Smith},
  {Weiland}, {Gold}, {Jarosik}, {Kogut}, {Limon}, {Meyer}, {Tucker}, {Wollack},
  \& {Wright}}]{2013ApJS..208...19H}
{Hinshaw}, G., {Larson}, D., {Komatsu}, E., {et~al.} 2013, \apjs, 208, 19

\bibitem[{{Hoeft} {et~al.}(2006){Hoeft}, {Yepes}, {Gottl{\"o}ber}, \&
  {Springel}}]{2006MNRAS.371..401H}
{Hoeft}, M., {Yepes}, G., {Gottl{\"o}ber}, S., \& {Springel}, V. 2006, \mnras,
  371, 401

\bibitem[{{Hotchkiss} {et~al.}(2015){Hotchkiss}, {Nadathur}, {Gottl{\"o}ber},
  {Iliev}, {Knebe}, {Watson}, \& {Yepes}}]{2015MNRAS.446.1321H}
{Hotchkiss}, S., {Nadathur}, S., {Gottl{\"o}ber}, S., {et~al.} 2015, \mnras,
  446, 1321

\bibitem[{{Hoyle} {et~al.}(2005){Hoyle}, {Rojas}, {Vogeley}, \&
  {Brinkmann}}]{2005ApJ...620..618H}
{Hoyle}, F., {Rojas}, R.~R., {Vogeley}, M.~S., \& {Brinkmann}, J. 2005, \apj,
  620, 618

\bibitem[{{Hu} \& {Sugiyama}(1996)}]{1996ApJ...471..542H}
{Hu}, W., \& {Sugiyama}, N. 1996, \apj, 471, 542

\bibitem[{{Ili{\'c}} {et~al.}(2013){Ili{\'c}}, {Langer}, \&
  {Douspis}}]{2013A&A...556A..51I}
{Ili{\'c}}, S., {Langer}, M., \& {Douspis}, M. 2013, \aap, 556, A51

\bibitem[{{Inman} {et~al.}(2015){Inman}, {Emberson}, {Pen}, {Farchi}, {Yu}, \&
  {Harnois-D{\'e}raps}}]{2015PhRvD..92b3502I}
{Inman}, D., {Emberson}, J.~D., {Pen}, U.-L., {et~al.} 2015, \prd, 92, 023502

\bibitem[{{Inman} {et~al.}(2016){Inman}, {Yu}, {Zhu}, {Emberson}, {Pen},
  {Zhang}, {Yuan}, {Chen}, \& {Xing}}]{2016arXiv161009354I}
{Inman}, D., {Yu}, H.-R., {Zhu}, H.-M., {et~al.} 2016, ArXiv e-prints,
  arXiv:1610.09354

\bibitem[{{Jennings} {et~al.}(2013){Jennings}, {Li}, \&
  {Hu}}]{2013MNRAS.434.2167J}
{Jennings}, E., {Li}, Y., \& {Hu}, W. 2013, \mnras, 434, 2167

\bibitem[{{Kauffmann} \& {Fairall}(1991)}]{1991MNRAS.248..313K}
{Kauffmann}, G., \& {Fairall}, A.~P. 1991, \mnras, 248, 313

\bibitem[{{Kerscher} {et~al.}(2000){Kerscher}, {Szapudi}, \&
  {Szalay}}]{2000ApJ...535L..13K}
{Kerscher}, M., {Szapudi}, I., \& {Szalay}, A.~S. 2000, \apjl, 535, L13

\bibitem[{{Kitaura} {et~al.}(2015){Kitaura}, {Gil-Mar{\'{\i}}n},
  {Sc{\'o}ccola}, {Chuang}, {M{\"u}ller}, {Yepes}, \&
  {Prada}}]{2015MNRAS.450.1836K}
{Kitaura}, F.-S., {Gil-Mar{\'{\i}}n}, H., {Sc{\'o}ccola}, C.~G., {et~al.} 2015,
  \mnras, 450, 1836

\bibitem[{{Kitaura} {et~al.}(2014){Kitaura}, {Yepes}, \&
  {Prada}}]{2014MNRAS.439L..21K}
{Kitaura}, F.-S., {Yepes}, G., \& {Prada}, F. 2014, \mnras, 439, L21

\bibitem[{{Kitaura} {et~al.}(2016){Kitaura}, {Chuang}, {Liang}, {Zhao}, {Tao},
  {Rodr{\'{\i}}guez-Torres}, {Eisenstein}, {Gil-Mar{\'{\i}}n}, {Kneib},
  {McBride}, {Percival}, {Ross}, {S{\'a}nchez}, {Tinker}, {Tojeiro},
  {Vargas-Magana}, \& {Zhao}}]{2016PhRvL.116q1301K}
{Kitaura}, F.-S., {Chuang}, C.-H., {Liang}, Y., {et~al.} 2016, Physical Review
  Letters, 116, 171301

\bibitem[{{Komatsu} {et~al.}(2011){Komatsu}, {Smith}, {Dunkley}, {Bennett},
  {Gold}, {Hinshaw}, {Jarosik}, {Larson}, {Nolta}, {Page}, {Spergel},
  {Halpern}, {Hill}, {Kogut}, {Limon}, {Meyer}, {Odegard}, {Tucker}, {Weiland},
  {Wollack}, \& {Wright}}]{2011ApJS..192...18K}
{Komatsu}, E., {Smith}, K.~M., {Dunkley}, J., {et~al.} 2011, \apjs, 192, 18

\bibitem[{{Lam} {et~al.}(2015){Lam}, {Clampitt}, {Cai}, \&
  {Li}}]{2015MNRAS.450.3319L}
{Lam}, T.~Y., {Clampitt}, J., {Cai}, Y.-C., \& {Li}, B. 2015, \mnras, 450, 3319

\bibitem[{{Landy} \& {Szalay}(1993)}]{1993ApJ...412...64L}
{Landy}, S.~D., \& {Szalay}, A.~S. 1993, \apj, 412, 64

\bibitem[{{Laureijs}(2009)}]{2009arXiv0912.0914L}
{Laureijs}, R. 2009, ArXiv e-prints, arXiv:0912.0914

\bibitem[{{Lavaux} \& {Wandelt}(2010)}]{2010MNRAS.403.1392L}
{Lavaux}, G., \& {Wandelt}, B.~D. 2010, \mnras, 403, 1392

\bibitem[{{Lavaux} \& {Wandelt}(2012)}]{2012ApJ...754..109L}
---. 2012, \apj, 754, 109

\bibitem[{{Li}(2011)}]{2011MNRAS.411.2615L}
{Li}, B. 2011, \mnras, 411, 2615

\bibitem[{{Li} {et~al.}(2012){Li}, {Zhao}, \& {Koyama}}]{2012MNRAS.421.3481L}
{Li}, B., {Zhao}, G.-B., \& {Koyama}, K. 2012, \mnras, 421, 3481

\bibitem[{{Liang} {et~al.}(2016){Liang}, {Zhao}, {Chuang}, {Kitaura}, \&
  {Tao}}]{2016MNRAS.459.4020L}
{Liang}, Y., {Zhao}, C., {Chuang}, C.-H., {Kitaura}, F.-S., \& {Tao}, C. 2016,
  \mnras, 459, 4020

\bibitem[{{LSST Dark Energy Science Collaboration}(2012)}]{2012arXiv1211.0310L}
{LSST Dark Energy Science Collaboration}. 2012, ArXiv e-prints, arXiv:1211.0310

\bibitem[{{Martino} \& {Sheth}(2009)}]{2009arXiv0911.1829M}
{Martino}, M.~C., \& {Sheth}, R.~K. 2009, ArXiv e-prints, arXiv:0911.1829

\bibitem[{{Massara} {et~al.}(2015){Massara}, {Villaescusa-Navarro}, {Viel}, \&
  {Sutter}}]{2015JCAP...11..018M}
{Massara}, E., {Villaescusa-Navarro}, F., {Viel}, M., \& {Sutter}, P.~M. 2015,
  \jcap, 11, 018

\bibitem[{{Padilla} {et~al.}(2005){Padilla}, {Ceccarelli}, \&
  {Lambas}}]{2005MNRAS.363..977P}
{Padilla}, N.~D., {Ceccarelli}, L., \& {Lambas}, D.~G. 2005, \mnras, 363, 977

\bibitem[{{Park} \& {Lee}(2007)}]{2007PhRvL..98h1301P}
{Park}, D., \& {Lee}, J. 2007, Physical Review Letters, 98, 081301

\bibitem[{{Patiri} {et~al.}(2006{\natexlab{a}}){Patiri}, {Betancort-Rijo},
  {Prada}, {Klypin}, \& {Gottl{\"o}ber}}]{2006MNRAS.369..335P}
{Patiri}, S.~G., {Betancort-Rijo}, J.~E., {Prada}, F., {Klypin}, A., \&
  {Gottl{\"o}ber}, S. 2006{\natexlab{a}}, \mnras, 369, 335

\bibitem[{{Patiri} {et~al.}(2006{\natexlab{b}}){Patiri}, {Prada}, {Holtzman},
  {Klypin}, \& {Betancort-Rijo}}]{2006MNRAS.372.1710P}
{Patiri}, S.~G., {Prada}, F., {Holtzman}, J., {Klypin}, A., \&
  {Betancort-Rijo}, J. 2006{\natexlab{b}}, \mnras, 372, 1710

\bibitem[{{Peebles} \& {Hauser}(1974)}]{1974ApJS...28...19P}
{Peebles}, P.~J.~E., \& {Hauser}, M.~G. 1974, \apjs, 28, 19

\bibitem[{{Peebles} \& {Yu}(1970)}]{1970ApJ...162..815P}
{Peebles}, P.~J.~E., \& {Yu}, J.~T. 1970, \apj, 162, 815

\bibitem[{{Peloso} {et~al.}(2015){Peloso}, {Pietroni}, {Viel}, \&
  {Villaescusa-Navarro}}]{2015JCAP...07..001P}
{Peloso}, M., {Pietroni}, M., {Viel}, M., \& {Villaescusa-Navarro}, F. 2015,
  \jcap, 7, 001

\bibitem[{{Percival} {et~al.}(2007){Percival}, {Cole}, {Eisenstein}, {Nichol},
  {Peacock}, {Pope}, \& {Szalay}}]{2007MNRAS.381.1053P}
{Percival}, W.~J., {Cole}, S., {Eisenstein}, D.~J., {et~al.} 2007, \mnras, 381,
  1053

\bibitem[{{Percival} {et~al.}(2010){Percival}, {Reid}, {Eisenstein}, {Bahcall},
  {Budavari}, {Frieman}, {Fukugita}, {Gunn}, {Ivezi{\'c}}, {Knapp}, {Kron},
  {Loveday}, {Lupton}, {McKay}, {Meiksin}, {Nichol}, {Pope}, {Schlegel},
  {Schneider}, {Spergel}, {Stoughton}, {Strauss}, {Szalay}, {Tegmark},
  {Vogeley}, {Weinberg}, {York}, \& {Zehavi}}]{2010MNRAS.401.2148P}
{Percival}, W.~J., {Reid}, B.~A., {Eisenstein}, D.~J., {et~al.} 2010, \mnras,
  401, 2148

\bibitem[{{Pisani} {et~al.}(2015){Pisani}, {Sutter}, {Hamaus}, {Alizadeh},
  {Biswas}, {Wandelt}, \& {Hirata}}]{2015PhRvD..92h3531P}
{Pisani}, A., {Sutter}, P.~M., {Hamaus}, N., {et~al.} 2015, \prd, 92, 083531

\bibitem[{{Planck Collaboration} {et~al.}(2014){Planck Collaboration}, {Ade},
  {Aghanim}, {Alves}, {Armitage-Caplan}, {Arnaud}, {Ashdown},
  {Atrio-Barandela}, {Aumont}, {Aussel}, \& et~al.}]{2014A&A...571A...1P}
{Planck Collaboration}, {Ade}, P.~A.~R., {Aghanim}, N., {et~al.} 2014, \aap,
  571, A1

\bibitem[{{Politzer} \& {Preskill}(1986)}]{1986PhRvL..56...99P}
{Politzer}, H.~D., \& {Preskill}, J.~P. 1986, Physical Review Letters, 56, 99

\bibitem[{{Pons-Border{\'{\i}}a} {et~al.}(1999){Pons-Border{\'{\i}}a},
  {Mart{\'{\i}}nez}, {Stoyan}, {Stoyan}, \& {Saar}}]{1999ApJ...523..480P}
{Pons-Border{\'{\i}}a}, M.-J., {Mart{\'{\i}}nez}, V.~J., {Stoyan}, D.,
  {Stoyan}, H., \& {Saar}, E. 1999, \apj, 523, 480

\bibitem[{{Pustilnik} {et~al.}(2002){Pustilnik}, {Martin}, {Huchtmeier},
  {Brosch}, {Lipovetsky}, \& {Richter}}]{2002A&A...389..405P}
{Pustilnik}, S.~A., {Martin}, J.-M., {Huchtmeier}, W.~K., {et~al.} 2002, \aap,
  389, 405

\bibitem[{{Pustilnik} {et~al.}(2011){Pustilnik}, {Martin}, {Tepliakova}, \&
  {Kniazev}}]{2011MNRAS.417.1335P}
{Pustilnik}, S.~A., {Martin}, J.-M., {Tepliakova}, A.~L., \& {Kniazev}, A.~Y.
  2011, \mnras, 417, 1335

\bibitem[{{Pycke} \& {Russell}(2016)}]{2016ApJ...821..110P}
{Pycke}, J.-R., \& {Russell}, E. 2016, \apj, 821, 110

\bibitem[{{Rasmussen} \& {Williams}(2006)}]{2006gpml.book.....R}
{Rasmussen}, C.~E., \& {Williams}, C.~K.~I. 2006, {Gaussian Processes for
  Machine Learning}

\bibitem[{{Reid} {et~al.}(2012){Reid}, {Samushia}, {White}, {Percival},
  {Manera}, {Padmanabhan}, {Ross}, {S{\'a}nchez}, {Bailey}, {Bizyaev},
  {Bolton}, {Brewington}, {Brinkmann}, {Brownstein}, {Cuesta}, {Eisenstein},
  {Gunn}, {Honscheid}, {Malanushenko}, {Malanushenko}, {Maraston}, {McBride},
  {Muna}, {Nichol}, {Oravetz}, {Pan}, {de Putter}, {Roe}, {Ross}, {Schlegel},
  {Schneider}, {Seo}, {Shelden}, {Sheldon}, {Simmons}, {Skibba}, {Snedden},
  {Swanson}, {Thomas}, {Tinker}, {Tojeiro}, {Verde}, {Wake}, {Weaver},
  {Weinberg}, {Zehavi}, \& {Zhao}}]{2012MNRAS.426.2719R}
{Reid}, B.~A., {Samushia}, L., {White}, M., {et~al.} 2012, \mnras, 426, 2719

\bibitem[{{Ross} {et~al.}(2015){Ross}, {Samushia}, {Howlett}, {Percival},
  {Burden}, \& {Manera}}]{2015MNRAS.449..835R}
{Ross}, A.~J., {Samushia}, L., {Howlett}, C., {et~al.} 2015, \mnras, 449, 835

\bibitem[{{Russell} \& {Pycke}(2017)}]{2017ApJ...835...69R}
{Russell}, E., \& {Pycke}, J.-R. 2017, \apj, 835, 69

\bibitem[{{Schaap} \& {van de Weygaert}(2000)}]{2000A&A...363L..29S}
{Schaap}, W.~E., \& {van de Weygaert}, R. 2000, \aap, 363, L29

\bibitem[{{Schlegel} {et~al.}(2011){Schlegel}, {Abdalla}, {Abraham}, {Ahn},
  {Allende Prieto}, {Annis}, {Aubourg}, {Azzaro}, {Baltay}, {Baugh}, {Bebek},
  {Becerril}, {Blanton}, {Bolton}, {Bromley}, {Cahn}, {Carton},
  {Cervantes-Cota}, {Chu}, {Cortes}, {Dawson}, {Dey}, {Dickinson}, {Diehl},
  {Doel}, {Ealet}, {Edelstein}, {Eppelle}, {Escoffier}, {Evrard}, {Faccioli},
  {Frenk}, {Geha}, {Gerdes}, {Gondolo}, {Gonzalez-Arroyo}, {Grossan},
  {Heckman}, {Heetderks}, {Ho}, {Honscheid}, {Huterer}, {Ilbert}, {Ivans},
  {Jelinsky}, {Jing}, {Joyce}, {Kennedy}, {Kent}, {Kieda}, {Kim}, {Kim},
  {Kneib}, {Kong}, {Kosowsky}, {Krishnan}, {Lahav}, {Lampton}, {LeBohec}, {Le
  Brun}, {Levi}, {Li}, {Liang}, {Lim}, {Lin}, {Linder}, {Lorenzon}, {de la
  Macorra}, {Magneville}, {Malina}, {Marinoni}, {Martinez}, {Majewski},
  {Matheson}, {McCloskey}, {McDonald}, {McKay}, {McMahon}, {Menard},
  {Miralda-Escude}, {Modjaz}, {Montero-Dorta}, {Morales}, {Mostek}, {Newman},
  {Nichol}, {Nugent}, {Olsen}, {Padmanabhan}, {Palanque-Delabrouille}, {Park},
  {Peacock}, {Percival}, {Perlmutter}, {Peroux}, {Petitjean}, {Prada},
  {Prieto}, {Prochaska}, {Reil}, {Rockosi}, {Roe}, {Rollinde}, {Roodman},
  {Ross}, {Rudnick}, {Ruhlmann-Kleider}, {Sanchez}, {Sawyer}, {Schimd},
  {Schubnell}, {Scoccimaro}, {Seljak}, {Seo}, {Sheldon}, {Sholl},
  {Shulte-Ladbeck}, {Slosar}, {Smith}, {Smoot}, {Springer}, {Stril}, {Szalay},
  {Tao}, {Tarle}, {Taylor}, {Tilquin}, {Tinker}, {Valdes}, {Wang}, {Wang},
  {Weaver}, {Weinberg}, {White}, {Wood-Vasey}, {Yang}, {Yeche}, {Zakamska},
  {Zentner}, {Zhai}, \& {Zhang}}]{2011arXiv1106.1706S}
{Schlegel}, D., {Abdalla}, F., {Abraham}, T., {et~al.} 2011, ArXiv e-prints,
  arXiv:1106.1706

\bibitem[{{Seo} \& {Eisenstein}(2003)}]{2003ApJ...598..720S}
{Seo}, H.-J., \& {Eisenstein}, D.~J. 2003, \apj, 598, 720

\bibitem[{{Shandarin} {et~al.}(2006){Shandarin}, {Feldman}, {Heitmann}, \&
  {Habib}}]{2006MNRAS.367.1629S}
{Shandarin}, S., {Feldman}, H.~A., {Heitmann}, K., \& {Habib}, S. 2006, \mnras,
  367, 1629

\bibitem[{{Sheth} \& {van de Weygaert}(2004)}]{2004MNRAS.350..517S}
{Sheth}, R.~K., \& {van de Weygaert}, R. 2004, \mnras, 350, 517

\bibitem[{{Slosar} {et~al.}(2013){Slosar}, {Ir{\v s}i{\v c}}, {Kirkby},
  {Bailey}, {Busca}, {Delubac}, {Rich}, {Aubourg}, {Bautista}, {Bhardwaj},
  {Blomqvist}, {Bolton}, {Bovy}, {Brownstein}, {Carithers}, {Croft}, {Dawson},
  {Font-Ribera}, {Le Goff}, {Ho}, {Honscheid}, {Lee}, {Margala}, {McDonald},
  {Medolin}, {Miralda-Escud{\'e}}, {Myers}, {Nichol}, {Noterdaeme},
  {Palanque-Delabrouille}, {P{\^a}ris}, {Petitjean}, {Pieri}, {Pi{\v s}kur},
  {Roe}, {Ross}, {Rossi}, {Schlegel}, {Schneider}, {Suzuki}, {Sheldon},
  {Seljak}, {Viel}, {Weinberg}, \& {Y{\`e}che}}]{2013JCAP...04..026S}
{Slosar}, A., {Ir{\v s}i{\v c}}, V., {Kirkby}, D., {et~al.} 2013, \jcap, 4, 026

\bibitem[{{Smoot} {et~al.}(1992){Smoot}, {Bennett}, {Kogut}, {Wright}, {Aymon},
  {Boggess}, {Cheng}, {de Amici}, {Gulkis}, {Hauser}, {Hinshaw}, {Jackson},
  {Janssen}, {Kaita}, {Kelsall}, {Keegstra}, {Lineweaver}, {Loewenstein},
  {Lubin}, {Mather}, {Meyer}, {Moseley}, {Murdock}, {Rokke}, {Silverberg},
  {Tenorio}, {Weiss}, \& {Wilkinson}}]{1992ApJ...396L...1S}
{Smoot}, G.~F., {Bennett}, C.~L., {Kogut}, A., {et~al.} 1992, \apjl, 396, L1

\bibitem[{{Spergel} {et~al.}(2003){Spergel}, {Verde}, {Peiris}, {Komatsu},
  {Nolta}, {Bennett}, {Halpern}, {Hinshaw}, {Jarosik}, {Kogut}, {Limon},
  {Meyer}, {Page}, {Tucker}, {Weiland}, {Wollack}, \&
  {Wright}}]{2003ApJS..148..175S}
{Spergel}, D.~N., {Verde}, L., {Peiris}, H.~V., {et~al.} 2003, \apjs, 148, 175

\bibitem[{{Spergel} {et~al.}(2007){Spergel}, {Bean}, {Dor{\'e}}, {Nolta},
  {Bennett}, {Dunkley}, {Hinshaw}, {Jarosik}, {Komatsu}, {Page}, {Peiris},
  {Verde}, {Halpern}, {Hill}, {Kogut}, {Limon}, {Meyer}, {Odegard}, {Tucker},
  {Weiland}, {Wollack}, \& {Wright}}]{2007ApJS..170..377S}
{Spergel}, D.~N., {Bean}, R., {Dor{\'e}}, O., {et~al.} 2007, \apjs, 170, 377

\bibitem[{{Sunyaev} \& {Zeldovich}(1970)}]{1970Ap&SS...7....3S}
{Sunyaev}, R.~A., \& {Zeldovich}, Y.~B. 1970, \apss, 7, 3

\bibitem[{{Sutter} {et~al.}(2012){Sutter}, {Lavaux}, {Wandelt}, \&
  {Weinberg}}]{2012ApJ...761..187S}
{Sutter}, P.~M., {Lavaux}, G., {Wandelt}, B.~D., \& {Weinberg}, D.~H. 2012,
  \apj, 761, 187

\bibitem[{{Sutter} {et~al.}(2014){Sutter}, {Pisani}, {Wandelt}, \&
  {Weinberg}}]{2014MNRAS.443.2983S}
{Sutter}, P.~M., {Pisani}, A., {Wandelt}, B.~D., \& {Weinberg}, D.~H. 2014,
  \mnras, 443, 2983

\bibitem[{{Trujillo} {et~al.}(2006){Trujillo}, {Carretero}, \&
  {Patiri}}]{2006ApJ...640L.111T}
{Trujillo}, I., {Carretero}, C., \& {Patiri}, S.~G. 2006, \apjl, 640, L111

\bibitem[{{Upadhye} {et~al.}(2014){Upadhye}, {Biswas}, {Pope}, {Heitmann},
  {Habib}, {Finkel}, \& {Frontiere}}]{2014PhRvD..89j3515U}
{Upadhye}, A., {Biswas}, R., {Pope}, A., {et~al.} 2014, \prd, 89, 103515

\bibitem[{{van de Weygaert} \& {Schaap}(2009)}]{2009LNP...665..291V}
{van de Weygaert}, R., \& {Schaap}, W. 2009, in Lecture Notes in Physics,
  Berlin Springer Verlag, Vol. 665, Data Analysis in Cosmology, ed. V.~J.
  {Mart{\'{\i}}nez}, E.~{Saar}, E.~{Mart{\'{\i}}nez-Gonz{\'a}lez}, \& M.-J.
  {Pons-Border{\'{\i}}a}, 291--413

\bibitem[{{van de Weygaert} \& {van Kampen}(1993)}]{1993MNRAS.263..481V}
{van de Weygaert}, R., \& {van Kampen}, E. 1993, \mnras, 263, 481

\bibitem[{{Varela} {et~al.}(2012){Varela}, {Betancort-Rijo}, {Trujillo}, \&
  {Ricciardelli}}]{2012ApJ...744...82V}
{Varela}, J., {Betancort-Rijo}, J., {Trujillo}, I., \& {Ricciardelli}, E. 2012,
  \apj, 744, 82

\bibitem[{{Vargas-Maga{\~n}a} {et~al.}(2013){Vargas-Maga{\~n}a}, {Bautista},
  {Hamilton}, {Busca}, {Aubourg}, {Labatie}, {Le Goff}, {Escoffier}, {Manera},
  {McBride}, {Schneider}, \& {Willmer}}]{2013A&A...554A.131V}
{Vargas-Maga{\~n}a}, M., {Bautista}, J.~E., {Hamilton}, J.-C., {et~al.} 2013,
  \aap, 554, A131

\bibitem[{{Wertheim}(1963)}]{1963PhRvL..10..321W}
{Wertheim}, M.~S. 1963, Physical Review Letters, 10, 321

\bibitem[{{White} {et~al.}(2011){White}, {Blanton}, {Bolton}, {Schlegel},
  {Tinker}, {Berlind}, {da Costa}, {Kazin}, {Lin}, {Maia}, {McBride},
  {Padmanabhan}, {Parejko}, {Percival}, {Prada}, {Ramos}, {Sheldon}, {de
  Simoni}, {Skibba}, {Thomas}, {Wake}, {Zehavi}, {Zheng}, {Nichol},
  {Schneider}, {Strauss}, {Weaver}, \& {Weinberg}}]{2011ApJ...728..126W}
{White}, M., {Blanton}, M., {Bolton}, A., {et~al.} 2011, \apj, 728, 126

\bibitem[{{White}(1979)}]{1979MNRAS.186..145W}
{White}, S.~D.~M. 1979, \mnras, 186, 145

\bibitem[{{Xu} {et~al.}(2015){Xu}, {Wang}, \& {Chen}}]{2015ApJ...798...40X}
{Xu}, Y., {Wang}, X., \& {Chen}, X. 2015, \apj, 798, 40

\bibitem[{{York} {et~al.}(2000){York}, {Adelman}, {Anderson}, {Anderson},
  {Annis}, {Bahcall}, {Bakken}, {Barkhouser}, {Bastian}, {Berman}, {Boroski},
  {Bracker}, {Briegel}, {Briggs}, {Brinkmann}, {Brunner}, {Burles}, {Carey},
  {Carr}, {Castander}, {Chen}, {Colestock}, {Connolly}, {Crocker}, {Csabai},
  {Czarapata}, {Davis}, {Doi}, {Dombeck}, {Eisenstein}, {Ellman}, {Elms},
  {Evans}, {Fan}, {Federwitz}, {Fiscelli}, {Friedman}, {Frieman}, {Fukugita},
  {Gillespie}, {Gunn}, {Gurbani}, {de Haas}, {Haldeman}, {Harris}, {Hayes},
  {Heckman}, {Hennessy}, {Hindsley}, {Holm}, {Holmgren}, {Huang}, {Hull},
  {Husby}, {Ichikawa}, {Ichikawa}, {Ivezi{\'c}}, {Kent}, {Kim}, {Kinney},
  {Klaene}, {Kleinman}, {Kleinman}, {Knapp}, {Korienek}, {Kron}, {Kunszt},
  {Lamb}, {Lee}, {Leger}, {Limmongkol}, {Lindenmeyer}, {Long}, {Loomis},
  {Loveday}, {Lucinio}, {Lupton}, {MacKinnon}, {Mannery}, {Mantsch}, {Margon},
  {McGehee}, {McKay}, {Meiksin}, {Merelli}, {Monet}, {Munn}, {Narayanan},
  {Nash}, {Neilsen}, {Neswold}, {Newberg}, {Nichol}, {Nicinski}, {Nonino},
  {Okada}, {Okamura}, {Ostriker}, {Owen}, {Pauls}, {Peoples}, {Peterson},
  {Petravick}, {Pier}, {Pope}, {Pordes}, {Prosapio}, {Rechenmacher}, {Quinn},
  {Richards}, {Richmond}, {Rivetta}, {Rockosi}, {Ruthmansdorfer}, {Sandford},
  {Schlegel}, {Schneider}, {Sekiguchi}, {Sergey}, {Shimasaku}, {Siegmund},
  {Smee}, {Smith}, {Snedden}, {Stone}, {Stoughton}, {Strauss}, {Stubbs},
  {SubbaRao}, {Szalay}, {Szapudi}, {Szokoly}, {Thakar}, {Tremonti}, {Tucker},
  {Uomoto}, {Vanden Berk}, {Vogeley}, {Waddell}, {Wang}, {Watanabe},
  {Weinberg}, {Yanny}, {Yasuda}, \& {SDSS Collaboration}}]{2000AJ....120.1579Y}
{York}, D.~G., {Adelman}, J., {Anderson}, Jr., J.~E., {et~al.} 2000, \aj, 120,
  1579

\bibitem[{{Yu} {et~al.}(2017){Yu}, {Emberson}, {Inman}, {Zhang}, {Pen},
  {Harnois-D{\'e}raps}, {Yuan}, {Teng}, {Zhu}, {Chen}, {Xing}, {Du}, {Zhang},
  {Lu}, \& {Liao}}]{2017NatAs...1E.143Y}
{Yu}, H.-R., {Emberson}, J.~D., {Inman}, D., {et~al.} 2017, Nature Astronomy,
  1, 0143

\bibitem[{{Zhao} {et~al.}(2016){Zhao}, {Tao}, {Liang}, {Kitaura}, \&
  {Chuang}}]{2016MNRAS.459.2670Z}
{Zhao}, C., {Tao}, C., {Liang}, Y., {Kitaura}, F.-S., \& {Chuang}, C.-H. 2016,
  \mnras, 459, 2670

\bibitem[{{Zivick} {et~al.}(2015){Zivick}, {Sutter}, {Wandelt}, {Li}, \&
  {Lam}}]{2015MNRAS.451.4215Z}
{Zivick}, P., {Sutter}, P.~M., {Wandelt}, B.~D., {Li}, B., \& {Lam}, T.~Y.
  2015, \mnras, 451, 4215

\end{thebibliography}

\clearpage

\end{document}